\documentclass[aip,amsmath,amssymb,reprint,]{revtex4-1}

\usepackage{graphicx}
\usepackage{dcolumn}
\usepackage{bm}

\usepackage[utf8]{inputenc}
\usepackage[T1]{fontenc}
\usepackage{mathptmx}

\begin{document}

\title{Reconstruction and interpretation of photon Doppler velocimetry spectrum for ejecta particles from shock-loaded sample in vacuum}

\author{Xiao-Feng Shi}
\affiliation{Institute of Applied Physical and Computational Mathematics, Beijing 100094, China}

\author{Dong-Jun Ma}
\email{ma\_dongjun@iapcm.ac.cn}
\affiliation{Institute of Applied Physical and Computational Mathematics, Beijing 100094, China}

\author{Song-lin Dang}
\affiliation{Jiangxi University of Applied Science, Nanchang 330103, China}

\author{Zong-Qiang Ma}
\affiliation{Institute of Applied Physical and Computational Mathematics, Beijing 100094, China}

\author{Hai-Quan Sun}
\affiliation{Institute of Applied Physical and Computational Mathematics, Beijing 100094, China}

\author{An-Min He}
\affiliation{Institute of Applied Physical and Computational Mathematics, Beijing 100094, China}

\author{Pei Wang}
\email{wangpei@iapcm.ac.cn}
\affiliation{Institute of Applied Physical and Computational Mathematics, Beijing 100094, China}
\affiliation{Center for Applied Physics and Technology, Peking University, Beijing 100871, China}

\date{\today}

\begin{abstract}
The photon Doppler velocimetry (PDV) spectrum is investigated in an attempt to reveal the particle parameters of ejecta from shock-loaded samples in a vacuum. A GPU-accelerated Monte-Carlo algorithm, which considers the multiple-scattering effects of light, is applied to reconstruct the light field of the ejecta and simulate the corresponding PDV spectrum. The influence of the velocity profile, total area mass, and particle size of the ejecta on the simulated spectra is discussed qualitatively. To facilitate a quantitative discussion, a novel theoretical optical model is proposed in which the single-scattering assumption is applied. With this model, the relationships between the particle parameters of ejecta and the peak information of the PDV spectrum are derived, enabling direct extraction of the particle parameters from the PDV spectrum. The values of the ejecta parameters estimated from the experimental spectrum are in good agreement with those measured by a piezoelectric probe.
\end{abstract}

\maketitle

\section{Introduction}
The strong shock wave released from the metal--vacuum/gas interface may eject a great number of metal particles.\cite{Sollier:1,Monfared:1,Asay:1,Speight:1,Ogorodnikov:1} Most of these particles are of micrometer-scale in size. This phenomenon of ejecta, or microjetting, was first observed by Kormer et al. in a plane impact experiment in the 1950s.\cite{Ogorodnikov:1} And the earliest available technical report on ejecta is from research by the Atomic Weapons Research Establishment, Aldermaston(UK).\cite{Bistow:1} The physics of ejecta are understood as a special limiting case of impulse driven Richtmyer–Meshkov.\cite{Richtmyer:1, Meshkov:1} In recent decades, extensive investigations on particle ejection have been performed because of its important role in many scientific and engineering fields, including explosion damage,\cite{Yeager:1} pyrotechnics,\cite{Held:1} and inertial confinement fusion.\cite{Tokheim:1,Masters:1} Many experimental approaches have attempted to measure the ejection production, such as the Asay foil,\cite{Asay:1,Asay:2} foam recovery,\cite{He:1} piezoelectric probes,\cite{Speight:1,Vogan:1} Fraunhofer holography,\cite{Sorenson:1,Sorenson:2} X-ray/proton radiography,\cite{Monfared:1,Hammerberg:1} Mie scattering,\cite{Monfared:2,Hammerberg:1}, and photon Doppler velocimetry (PDV).\cite{La:1, Ogorodnikov:2, Andriyash:1, Franzkowiak:1, Sun:1, Fedorov:1, Arsenii:1} The main quantities of interest are the particles' velocity, diameter, and total area mass. Most approaches can only measure some of these ejecta parameters. To reveal the full particle field of ejecta, multiple measurement approaches must be equipped. However, in real-world conditions, these approaches are hard to apply simultaneously because of limitations on the experimental space or configuration. Recently, PDV has attracted considerable attention\cite{Andriyash:1, Franzkowiak:1, Arsenii:1} owing to its ability to recover the total area mass and the distributions of particle velocity and diameter at the same time. In addition, the light path of PDV is rather concise and its application is convenient. In some complex experimental configurations, PDV may be the only approach that can measure the ejecta particles.

\begin{figure}
\centering
\includegraphics[width=0.38\textwidth]{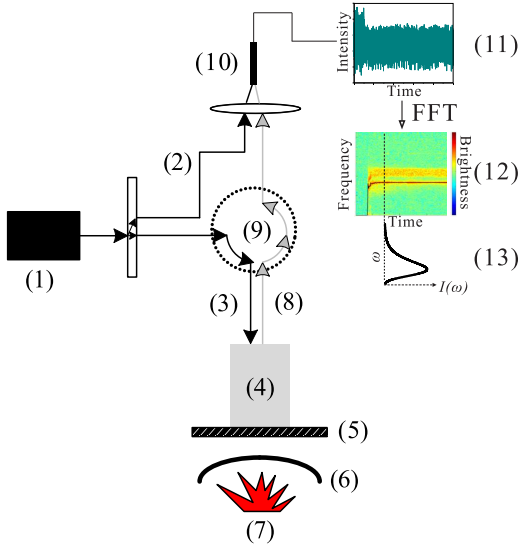}
\caption{ Standard PDV setup. (1) Laser; (2) Reference light; (3) Incident light; (4) Ejecta; (5) Metal plate; (6) Shock; (7) Detonation; (8) Backscattering light; (9) Optical circulator; (10) Photodetector; (11) Photoelectric signal; (12) PDV spectrogram; (13) Instantaneous spectrum.}
\label{fig1}
\end{figure}

A standard PDV setup is shown in Fig.~\ref{fig1}. The photodetector records a mixture of reference and backscattering light. The reference wave is in the carrier frequency, and the backscattering wave from ejecta particles has a shifted frequency due to the Doppler effect. The interference of the two light waves in the photodetector leads to temporal beats of light intensity. The beat signal consists of a large number of harmonics with different amplitudes and phases. The heterodyne signal may change according to variations in the particles' position and velocity. A discrete Fourier transform is applied to sweep the beats over time, giving a two-dimensional spectrogram on the ``frequency/velocity--time'' plane. In the spectrogram, the brightness of each point represents the corresponding spectral amplitude. The spectrogram is composed of the integral of all particles' scattering effects. Hence, interpreting the spectrogram in detail remains a challenging task.

There have been several studies on the interpretation of the PDV spectrum. Buttler\cite{Buttler:1,Buttler:2} used the spectrogram boundaries to determine the velocities of the spike and bubble of Richtmyer--Meshkov instability in loaded metal surface. The evolution of the PDV spectrogram in a gas environment was discussed by Sun et al.,\cite{Sun:1} and the upper boundary of the spectrogram was used to obtain the particle size by considering aerodynamic deceleration effects. Fedorov et al.\cite{Fedorov:1} discussed the influence of different particle sizes on the spectrogram boundary in further detail. Recently, Franzkowiak et al.\cite{Franzkowiak:1} and Andriyash et al.\cite{Andriyash:1,Arsenii:1} reconstructed the light field of ejecta and obtained the simulated PDV spectrum using single- and multiple-scattering theory, respectively. They varied the particles' parameters and fitted the simulated PDV spectrum to the experimental data. In this way, the particle velocity profile, diameter, and total area mass were recovered. Andriyash et al. considered the aerodynamic deceleration effects in a gas environment, whereas Franzkowiak et al. only discussed the case of a vacuum.

Franzkowaik et al.\cite{Franzkowiak:1} and Andriyash et al.\cite{Andriyash:1,Arsenii:1} proposed similar approaches for recovering the ejecta parameters from the PDV spectrum through reconstruction and then fitting. However, some assumptions were introduced in the reconstruction of the light field. Franzkowaik et al. assumed that only backscattering light was present, while Andriyash et al. set the light scattering direction to be uniform and random in space. These assumptions affect the accuracy of the spectrum reconstruction, and thus influence the recovery of the ejecta parameters. The fitting model is another factor that affects the interpretation of the PDV spectrum. Different convergence criteria may produce different results. The quantitative relationship between the ejecta parameters and the PDV spectrum remains unclear. Hence, it is difficult to obtain definite ejecta parameters from the PDV spectrum. These issues provide the motivation for the present work.

In this study, we improve the reconstruction method of the ejecta light field, and propose a novel model for extracting the ejecta parameters directly from the PDV spectrum. Mie theory, which gives a rigorous mathematical solution to Maxwell's equations, is applied to calculate the light scattering effects, and a Monte-Carlo (MC) algorithm is used to describe the light transport process realistically. This reconstruction method provides a high-fidelity simulation for the PDV spectrum. The procedure is discussed in detail in Section II. The influence of the ejecta parameters on the PDV parameters is then explored through MC simulations in Section III.A. In Section III.B, we propose an optical model that reveals the relationships between the PDV spectrum characteristics and the ejecta parameters. With this model, the ejecta parameters can be extracted directly from the PDV spectrum, instead of fitting to experimental data. In Section III.C, the estimated ejecta parameters from an experimental PDV spectrum are verified against those measured by a piezoelectric probe. Finally, the conclusions to this study are presented in Section IV.

\section{Reconstruction of PDV spectrum}
\subsection{Theoretical background}
\begin{figure}
\centering
\includegraphics[width=0.4\textwidth]{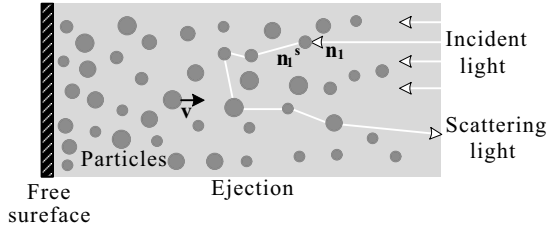}
\caption{Multiple scattering of light waves in ejection particles.}
\label{fig2}
\end{figure}

The photodetector records reference and backscattering light waves. The scattering process of incident light is illustrated in Fig.~\ref{fig2}. The scattering light from the ejecta is governed by the superposition of waves propagating in the ejection particles along different light paths \(i\):
\begin{equation}
{E_{bs}}\left( t \right) = \sum\limits_i {{E_i}\left( t \right)}
\label{eq1}
\end{equation}
where \(E_{bs}\) and \(E_{i}\) are the electric vectors of total and partial scattering waves, respectively.

The light intensity measured by the detector can be represented as:
\begin{equation}
\begin{array}{l}
I\left( t \right) = {\left( {{E_r}\left( t \right) + {E_{bs}}\left( t \right)} \right)^2}= {\left( {{E_r}\left( t \right) + \sum\limits_{i} {{E_i}\left( t \right)} } \right)^2}\\
= E_r^2\left( t \right) + \sum\limits_{i} {E_i^2\left( t \right)} + 2\sum\limits_{i} {{E_r}\left( t \right){E_i}\left( t \right) + } \sum\limits_{i \ne j} {{E_i}\left( t \right){E_j}\left( t \right)}
\end{array}
\label{eq2}
\end{equation}
where \(E_r^2\) and \(E_i^2\) denote the intensity of the reference and scattering light signals, respectively. The third term represents the heterodyne beats between the reference and scattering light, and the last term represents the heterodyne beats between the different scattering light paths. The Fourier transform of \(I(t)\) is determined by the relation:
\begin{equation}
\begin{array}{l}
I\left( \omega \right) = \int {dt\exp \left( {i\omega t} \right)} I\left( t \right)\\
\approx \int {dt\exp \left( {i\omega t} \right)} \left( {2\sum\limits_{i} {{E_r}\left( t \right){E_i}\left( t \right)} } \right)\\
= 2{\left| {{E_r}} \right|} \cdot {\left. {\sum\limits_{i} {\left| {{E_i}} \right|} } \right|_{\omega = {\omega _i} - {\omega _r}}}
\end{array}
\label{eq3}
\end{equation}
where \(\left| {{E}} \right|\) is the amplitude of the light wave field. In Eq.~(3), only the third term appearing in Eq.~(2) remains. This is because the frequencies of the first two terms are too high to be measured by the detector and the value of the last term is much smaller than that of the third term. \(\omega_r\) is the carrier frequency, and \(\omega_i\) is the Doppler-shifted frequency, which corresponds to a sequence of scattering events along path \(i\):
\begin{equation}
{\omega _i} = \frac{{{\omega _r}}}{c}\sum\limits_k {\left( {{\bf{n}}_{k,i}^s - {{\bf{n}}_{k,i}}} \right){{\bf{v}}_{k,i}}}
\label{eq4}
\end{equation}
where \(c\) is the speed of light, \({\bf{n}}_{k,i}\) and \({\bf{n}}_{k,i}^s\) are the directions of wave propagation before and after scattering by particle \(k\), and \({\bf{v}}_{k,i}\) is the velocity of this particle.

To reconstruct the PDV spectrogram [Eq.~(\ref{eq3})], the key is to obtain \(\left| {{E_i}} \right|\) and \(\omega_i\), i.e., the detailed scattering process in the particles. For multiple-particle systems, the scattering field can be described by the transport equation:\cite{Ishimaru:1, Reguigui:1, Binzoni:1}
\begin{equation}
\begin{split}
& \left( {\bf{n}\frac{\partial }{{\partial r}} + \sigma_s + \kappa } \right)I\left( {{\bf r,n},t} \right) \\
& = \int {\left\langle {\sigma_s p\left( {\bf n,n'} \right)\exp \left( {i{k_0}\left( {\bf n - n'} \right){\bf v}t} \right)} \right\rangle I\left( {{\bf r,n'},t} \right)d \bf n'}
\end{split}
\label{eq5}
\end{equation}
where \(I\) is the light intensity in the field, which depends on both the detection position \(\bf r\) and the direction \(\bf n\). \(\sigma_s\) and \(\kappa\) are the coefficients of scattering and absorption, respectively. \(p\left( {\bf n,n'} \right)\) is the scattering phase function. The right-hand side of this equation represents the contributions of scattering light from other positions.

At the boundaries of the ejection, the light intensity has the form:
\begin{equation}
\left\{ {\begin{array}{*{20}{l}}
{I\left( {{\bf{r}} = 0,{\bf{n}} = {{\bf{n}}_{\bf{0}}},t} \right) = {I_0}}\\
{I\left( {{\bf{r}} = 0,{\bf{n}} = - {{\bf{n}}_{\bf{0}}},t} \right) = {I_{bs}} = {{(\sum\limits_i {\left| {{E_i}} \right|} } )^2}}
\end{array}} \right.
\label{eq6}
\end{equation}
where \(\bf n_0\) is the direction of incident light, which is usually perpendicular to the free surface. \(I_0\) and \(I_{bs}\) are the intensities of incident light and backscattering light, respectively, which correspond to the input and output of the transport equation.

\subsection{Monte-Carlo algorithm}
Andriyash et al.\cite{Andriyash:1,Arsenii:1} used the discrete ordinate method to solve the transport equation [Eq.~(\ref{eq5})]. In this paper, a more convenient and accurate method of the MC algorithm is applied to calculate the scattering effects.

In the MC algorithm, the incident light is assumed to be a great number of photons. When passing through random granular media, only part of the photons can penetrate. The proportion of permeable photons is approximated by Beer--Lambert's law:\cite{Bohren:1}
\begin{equation}
p_r =\exp \left( { - \tau L} \right)
\label{eq7}
\end{equation}
where \(L\) is the thickness of the medium and \(\tau \) denotes the inverse extinction length, given by:
\begin{equation}
\tau = N{K_{ext}}\bar A = \frac{{\sum\limits_i {\frac{\pi }{4}{d_i}^2 \cdot {K_{ext}}\left( {{d_i}} \right)} }}{{{A_s}L}}
\label{eq8}
\end{equation}
where \(N \) is the number of particles per unit volume, \(K_{ext} \) is the light extinction coefficient (determined by the particle diameter, light wavelength, and metal relative refraction index), \(\bar A\) is the mean cross-section area of the particles, \(d_i\) is the diameter of particle \(i\), and \(A_s\) is the light exposure area.

For particles in the light exposure area, the total mass has the form:
\begin{equation}
{m_0}{A_s} = \sum\limits_i {\frac{1}{6}\pi {d_i^3}{\rho _0}}
\label{eq9}
\end{equation}
where \(m_0\) is the total area mass of ejection and \(\rho_0\) is the density of the particle material.

Combining Eqs.~(\ref{eq8}) and (\ref{eq9}), we can rewrite Eq.~(\ref{eq7}) as:
\begin{equation}
{p_r} = \exp \left( { - \frac{{3{m_0}\sum\limits_i {d_i^2{K_{ext}}\left( {{d_i}} \right)} }}{{2{\rho _0}\sum\limits_i {d_i^3} }}} \right)
\label{eq10}
\end{equation}

The photons staying in the medium are scattered or absorbed by particles. The probabilities of scattering and absorption are calculated by the formula:
\begin{equation}
\left\{ \begin{array}{l}
{p_s} = {K_{sca}}/{K_{ext}}\\
{p_a} = {K_{abs}}/{K_{ext}}
\end{array} \right.
\label{eq11}
\end{equation}
where the scattering coefficient \(K_{sca}\) and the absorption coefficient \(K_{abs}\) are calculated by Mie theory\cite{Mie:1} as:
\begin{equation}
\left\{ \begin{array}{l}
{K_{ext}} = \frac{2}{{{\alpha ^2}}}\sum\limits_{n = 1}^\infty {\left( {2n + 1} \right){\mathop{\rm Re}\nolimits} \left( {{a_n} + {b_n}} \right)} \\
{K_{sca}} = \frac{2}{{{\alpha ^2}}}\sum\limits_{n = 1}^\infty {\left( {2n + 1} \right)\left( {{{\left| {{a_n}} \right|}^2} + {{\left| {{b_n}} \right|}^2}} \right)} \\
{K_{abs}} = {K_{ext}} - {K_{sca}}
\end{array} \right.
\label{eq12}
\end{equation}
where \(\alpha\) is a dimensionless particle diameter parameter, \(\alpha = \pi d/\lambda \), \(\lambda\) is the light wavelength, and \(a_n\), \(b_n\) are Mie coefficients. It is clear that \(p_s+p_a=1\).

If the photon is absorbed by particles, it will completely disappear and be converted into the particle's internal energy. If the photon is scattered, its propagation direction and frequency will change, as shown in Fig.~3. The phase function of the scattering angle \(\theta\) is calculated by the formula:
\begin{equation}
p(\theta ) = \frac{{{\lambda ^2}}}{{2\pi {K_{sca}}}}({\left| {{S_1}(\theta )} \right|^2} + {\left| {{S_2}(\theta )} \right|^2})
\label{eq13}
\end{equation}
where \(S_1\) and \(S_2\) denote the scattering intensity components in the perpendicular and parallel directions:
\begin{equation}
\left\{ \begin{array}{l}
{S_1}\left( \theta \right) = \sum\limits_{n = 1}^\infty {\frac{{2n + 1}}{{n\left( {n + 1} \right)}}\left( {{a_n}\frac{{d{P_n}\left( {\cos \theta } \right)}}{{d\cos \theta }} + {b_n}\frac{{dP_n^{(1)}\left( {\cos \theta } \right)}}{{d\theta }}} \right)} \\
{S_2}\left( \theta \right) = \sum\limits_{n = 1}^\infty {\frac{{2n + 1}}{{n\left( {n + 1} \right)}}\left( {{a_n}\frac{{dP_n^{(1)}\left( {\cos \theta } \right)}}{{d\theta }} + {b_n}\frac{{d{P_n}\left( {\cos \theta } \right)}}{{d\cos \theta }}} \right)}
\end{array} \right.
\label{eq14}
\end{equation}
where \(P_n\) and \(P_n^{(1)}\) are Legendre and first-order associative Legendre functions, respectively.

The incident light is assumed to be non-polarized, so the azimuth angle \(\varphi\) after scattering obeys the uniform random distribution:
\begin{equation}
p(\varphi)=\frac{1}{2\pi}
\label{eq15}
\end{equation}

\begin{figure}
\centering
\includegraphics[width=0.30\textwidth]{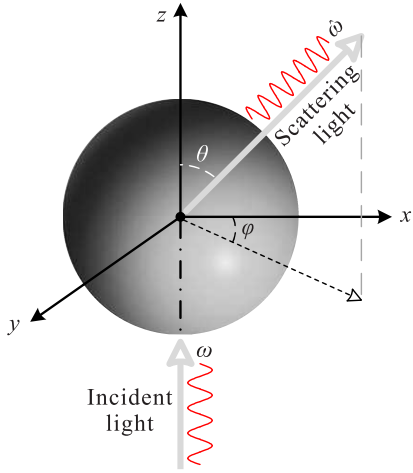}
\caption{Light scattering on a particle.}
\label{fig3}
\end{figure}

After scattering, the scattering angle and azimuth angle are added to the original light direction. The new direction cosine \(\hat u = [\hat u_x, \hat u_y, \hat u_z] \) has the form:
\begin{equation}
\left\{ \begin{array}{l}
{{\hat u}_x} = \frac{1}{{\sqrt {1 - u_z^2} }}\sin \left( \theta \right)\left[ {{u_x}{u_y}\cos \left( \varphi \right) - {u_y}\sin \left( \varphi \right)} \right] + {u_x}\cos \left( \theta \right)\\
{{\hat u}_y} = \frac{1}{{\sqrt {1 - u_z^2} }}\sin \left( \theta \right)\left[ {{u_x}{u_z}\cos \left( \varphi \right) - {u_x}\sin \left( \varphi \right)} \right] + {u_y}\cos \left( \theta \right)\\
{{\hat u}_z} = - \sin \left( \theta \right)\cos (\varphi )\sqrt {1 - u_z^2} + {u_z}\cos \left( \theta \right)
\end{array} \right.
\label{eq16}
\end{equation}
where \(u = [u_x, u_y, u_z] \) is the original direction cosine. When \(\left| {{u_z}} \right| \approx 1 \), the direction cosine \(\hat u\) is calculated by the formula:
\begin{equation}
\left\{ \begin{array}{l}
{{\hat u}_x} = \sin \left( \theta \right)\cos \left( \varphi \right)\\
{{\hat u}_y} = \sin \left( \theta \right)\sin \left( \varphi \right)\\
{{\hat u}_z} = \cos \left( \theta \right){u_z}/\left| {{u_z}} \right|
\end{array} \right.
\label{eq17}
\end{equation}

After scattering, the frequency of the photon will have changed. The new frequency of scattering light has the form:
\begin{equation}
\hat \omega = \omega \left( {1 + \frac{{{\bf{\hat v}} - {\bf{v}}}}{c}{\bf{n}}} \right)
\label{eq18}
\end{equation}
where \(\bf{v}\) is the velocity of the last particle that the photon left, \(\bf{\hat v}\) is the current particle velocity, \(\bf n\) is the photon direction, and \(\omega\) is the photon frequency before scattering.

After passing through the entire ejection layer, few photons reach the free surface. An ideal diffuse reflection is assumed for these photons. After reflection, the space angles \(\theta\) and \(\varphi\) are uniformly random in \([\pi/2, \pi]\) and \([0, 2\pi]\), respectively.

Because the ejection is blocked by the free surface, eventually all the photons are either absorbed by particles or backscattered out from the top of the ejection. The frequency shifts of these ``out'' photons are summarized as the theoretical PDV spectrogram:

\begin{equation}
I\left( \omega \right) \propto \sum\limits_i {\left| {{E_i}} \right|} = \sum\limits_i {\sqrt {{I_i}} } = {\left. {{n_{out}}} \right|_{\omega = \hat \omega - {\omega _r}}} \cdot \sqrt {\frac{{{I_0}}}{{{n_0}}}}
\label{eq19}
\end{equation}
where \(n_0\) is the number of initial photons and \(n_{out}\) is the distribution of out photons in terms of their frequency.

The detailed steps of the calculation procedure are as follows:

(1) First, the initial conditions of the photons and particles are set, such as the number and frequency of photons, and the diameter, velocity, and position of the particles. The photons start at the top of the ejection and then move towards the free surface.

(2) The step sizes of all photons are set to the same and equal to one thousandth of the height of the ejecta.

(3) In one iteration, all photons take one step in the direction of their propagation. Some photons may penetrate the current ejection layer, and the proportion \(p_r\) is determined by Eq.~(\ref{eq10}). For each photon, a random number is generated in (0, 1). If the random number is less then \(p_r\), the corresponding photon travels over the ejection layer boundary successfully. Otherwise, the corresponding photon is absorbed or scattered by particles in the ejection layer.

(4) For the photons that remain in the ejection layer, we use Eq.~(\ref{eq11}) to determine whether they are scattered or absorbed. If the photon is scattered, the direction change is calculated by Eqs.~(\ref{eq13})--(\ref{eq17}). Because the phase function of the scattering angle is very complex, an acceptance--rejection method is applied. The new frequency of the scattered photons is determined by Eq.~(\ref{eq18}).

(5) Overall, if the photon travels across the ejection layer boundary, its position is updated; if the photon is scattered, its direction, frequency, and position are updated; if the photon is absorbed, it is labeled as such and removed from subsequent calculations.

(6) After updating the state of the photons, we check which of them have reached the free surface or left through the top of the ejection. For all photons that have reached the free surface, the ideal diffuse reflection is applied. If any photons have left the ejection, they are labeled accordingly and removed from subsequent calculations.

(7) Steps (3)--(6) are repeated until all photons have been absorbed or have left the ejection. The frequency shifts of outgoing photons are summarized as the spectrogram.

\subsection{Particle models}
The MC algorithm indicates that the PDV spectrum is related to the particle size \(d\), velocity \(v\), position \(z\), and number \(N\) (i.e., total area mass \(m_0\)). This algorithm can be applied in cases where these parameters are completely random. In real situations, however, the particles of the ejecta usually satisfy certain distributions in terms of velocity and diameter.\cite{Monfared:1,Monfared:2,Durand:1,Durand:2,Schauer:1,Sorenson:1,Sorenson:2,He:2} For the sake of discussion, these assumptions are applied in this paper. Previous studies\cite{Monfared:1,Durand:1} indicate that the initial velocities of particles in the ejecta can be approximated by an exponential law:
\begin{equation}
f\left( v \right) = \frac{{m(v)}}{{{m_0}}} = \frac{\beta }{{{v_{fs}}}}\exp \left[ { - \beta \left( {\frac{v}{{{v_{fs}}}} - 1} \right)} \right]
\label{eq20}
\end{equation}
where \(v_{fs}\) is the velocity of the free surface and \(\beta\) is the velocity distribution coefficient. Under this exponential law, most of the particles are located in the low-velocity region, which is near the free surface. \(\beta\) determines the non-uniformity of this distribution.

In this paper, we only consider the ejecta in a vacuum environment. After being ejected, the particles retain an almost constant velocity and the ejecta expands in a self-similar manner over time. The particle position \(z\) is only related to its initial velocity \(v\) and ejection time \(t_e\), \(z=vt_e\). The corresponding PDV spectrum exhibits slight changes over time.\cite{Franzkowiak:1,Bell:1}

The particle size distribution is assumed to obey a log-normal law:\cite{Schauer:1,Sorenson:2}
\begin{equation}
n\left( d \right) = \frac{1}{{\sqrt {2\pi } \sigma d}}\exp \left( { - \frac{{{{\ln }^2}\left( {d/{d_m}} \right)}}{{2{\sigma ^2}}}} \right)
\label{eq21}
\end{equation}
where \(\sigma\) is the width of the distribution and \(d_m\) is the median diameter. These parameters depend on the roughness of the metal surface, shock-induced breakout pressure, and surrounding gas properties. When $\sigma$ = 0, the function becomes a Dirac equation and all of the particles have the same diameter. Obviously, this is the ideal situation. The particle distribution can also be described by a power law,\cite{Durand:2, Sorenson:1, He:2} but this description may be invalid in the range of small particle sizes (less than 10\; \(\rm \mu m\)).\cite{Sorenson:2} In this paper, the particle size is assumed to be independent of its velocity.

With these assumptions, the determining factors of the PDV spectrum change to the velocity profile coefficient \(\beta\), total area mass \(m_0\), median diameter \(d_m\), and size distribution width \(\sigma\). The aim of this paper is to discuss the influence of these parameters on the PDV spectrum, and to explore how they can be extracted from the PDV spectrum most accurately.

\subsection{Convergence and comparison}
\begin{figure}
\centering
\includegraphics[height=0.35\textwidth]{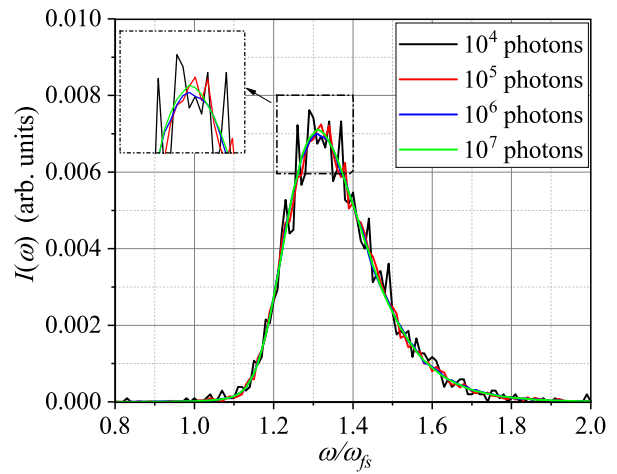}
\caption{Theoretical PDV spectra with different initial numbers of photons. The calculation assumes the ejection of Sn particles in a vacuum environment. The particle velocities obey an exponential distribution \((\beta=10)\) and the particle sizes follow a log-normal distribution \((d_m=1.5\; \rm{\mu m}, \sigma=0.5)\). The total area mass is \(20\; \rm{mg/cm^2}\). \(\omega_{fs}\) is the Doppler frequency shift corresponding to the free surface velocity, \(\omega_{fs}=2\omega_r \cdot v_{fs}/c\). The probing wavelength is \(\lambda=1550\; \rm{nm}\).}
\label{fig4}
\end{figure}

The accuracy of the MC algorithm mainly depends on the initial number of photons. Theoretical PDV spectra with \(10^4\), \(10^5\), \(10^6\), and \(10^7\) initial photons are shown in Fig.~\ref{fig4}. The calculation assumes a vacuum environment and the particles distribution assumptions are applied. In this case, the PDV spectra have a single peak. As the initial number of photons increases, the spectrum curves tend to be smooth. The difference between the spectra with \(10^6\) and \(10^7\) initial photons is very slight. Thus, \(10^7\) initial photons are applied in the following calculations.

\begin{figure}
\centering
\includegraphics[height=0.35\textwidth]{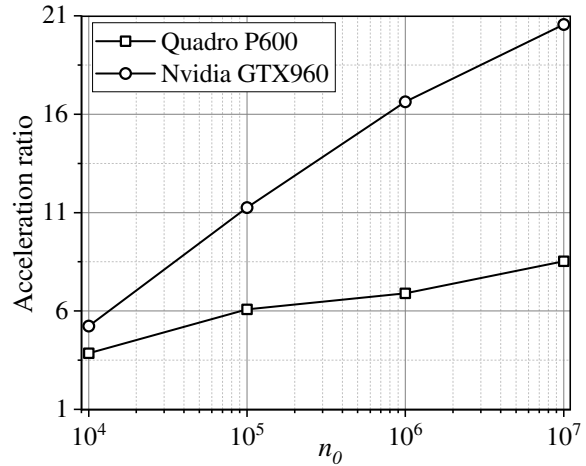}
\caption{Acceleration ratio of GPU calculation compared with CPU for different initial numbers of photons. The CPU is an Intel Xeon W-2102 and its basic frequency is 2.9 GHz. The Quadro P600 GPU has 384 stream processors; the clock speed of each processor is about 1.3 GHz. The Nvidia GTX960 GPU has 1024 stream processors; the clock speed of each processor is about 1.1 GHz.}
\label{fig5}
\end{figure}

The high initial number of photons leads to considerable computational cost. For the case of \(10^7\) photons, a single-core CPU requires approximately 3 h to determine the spectrum. GPUs can be applied to accelerate the calculation. Although the frequency of GPU processors is much lower than that of CPUs, GPUs contain hundreds or thousands of stream processors that can work simultaneously. The acceleration ratio of a GPU compared to a CPU is shown in Fig.~\ref{fig5}. The Intel Xeon W-2102 CPU (frequency 2.9 GHz) and two GPUs (Quadro P600 and Nvidia GTX960) are applied. As the initial number of photons increases, the acceleration ratio of the GPUs is enhanced. For the case of \(10^7\) photons, the acceleration ratio reaches a factor of 8 for the Quadro P600 and a factor of 20 for the Nvidia GTX960. Because there are many judgment events in the procedure, and the GPUs have few logical units, it is difficult to improve the acceleration ratio with these GPUs. Thus, the Nvidia GTX960, which requires approximately 500 s to compute each case, is used in the following calculations.

\begin{figure}
\centering
\includegraphics[height=0.37\textwidth]{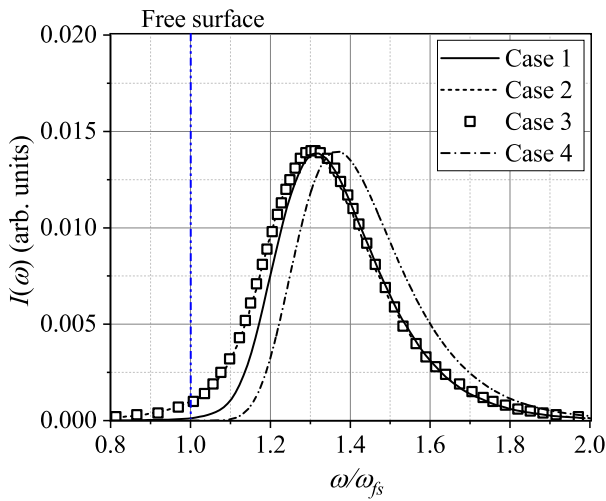}
\caption{Comparison of PDV spectra calculated by different reconstruction methods. Case 1: MC + Mie scattering theory (proposed procedure); Case 2: MC + uniform scattering assumption; Case 3: Discrete coordinates + uniform scattering assumption (Andriyash et al.); Case 4: Single scattering theory (Franzkowiak et al.) . The data for Case 3 were extracted directly from the paper of Andriyash et al. The calculations were carried out for a transport scattering thickness of \(\tau_{tr}=10\), which corresponds to \(m_0=10.7\; {\rm {mg/cm^2}}, d_m=1.5\; \rm {\mu m}\), and \(\sigma = 0.5\). The material is Sn and the ejection velocity profile has \(\beta=8\).}
\label{fig6}
\end{figure}

The PDV spectra simulated by the present procedure are compared with those reported by Andriyash et al. and Franzkowiak et al.  in Fig.~\ref{fig6}. We use the equivalent ejecta area mass and particle size instead of the transport optical thickness used by Andriyash et al. With the uniform scattering assumption, our simulation (Case 2) is almost the same as that of Andriyash et al. (Case 3), which validates the adequacy of the present numerical method. However, when the Mie scattering theory is applied, there is a remarkable difference between the present procedure (Case 1) and the results of Andriyash et al. (Case 3) and Franzkowiak et al. (Case 4). The difference with Andriyash et al. is mainly in the low-velocity part. This is because the change in the scattering phase function has a great influence on the multiple scattering, which is the main form of scattering in the low-velocity dense part. The difference with Franzkowiak et al. is in the location of the spectrum peak. Franzkowiak et al. applied the single-scattering theory and assumed that all of the light scattered backward. This implies that the optical thickness is overestimated, and so little light would reach the deep region of the ejecta. Thus, the spectrum moves toward high velocities. These differences in spectra indicate that the scattering assumption may introduce some reconstruction inaccuracy that cannot be neglected.

\section{Interpretation of PDV spectrum}
\subsection{Influence of ejecta parameters}
\begin{figure}
\centering
\includegraphics[height=0.35\textwidth]{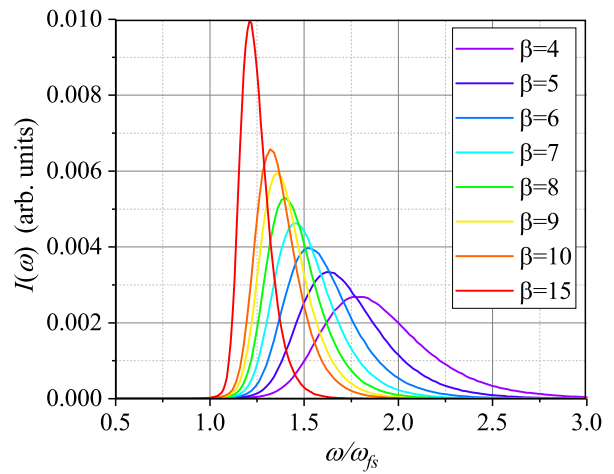}
\caption{Simulated PDV spectra with different velocity coefficients. The calculation was carried out for Sn particles in a vacuum environment. The total area mass was \(20\; \rm mg/cm^2\). The log-normal distribution \((d_m=1.5\; {\rm \mu m}, \sigma = 0.5)\) was applied to the particle sizes.}
\label{fig7}
\end{figure}

\begin{figure}
\centering
\includegraphics[height=0.35\textwidth]{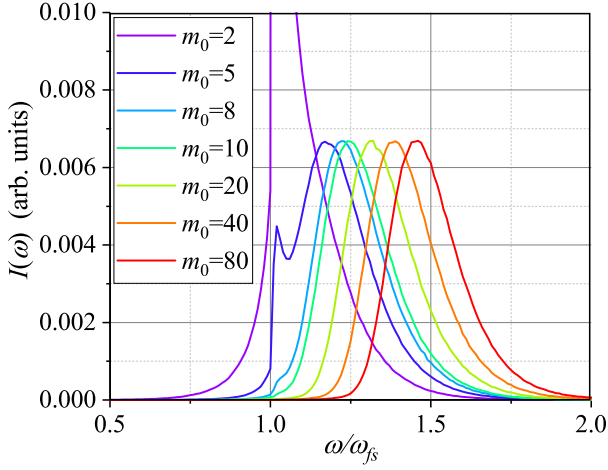}
\caption{Simulated PDV spectra with different total area mass. The area mass unit is \(\rm {mg/cm^2}\). The calculation was carried out for Sn particles in a vacuum environment. The velocity coefficient \(\beta = 10\) and the size coefficients \(d_m=1.5\; \rm {\mu m}, \sigma = 0.5\).}
\label{fig8}
\end{figure}

\begin{figure}
\centering
\includegraphics[height=0.35\textwidth]{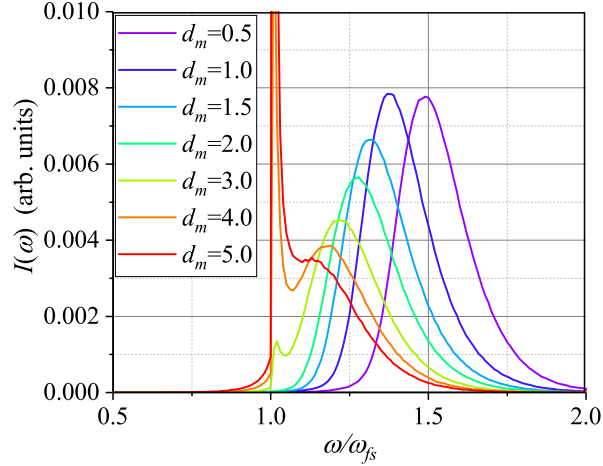}
\caption{Simulated PDV spectra with different particle median diameters. The diameter unit is \(\rm {\mu m}\). The calculation was carried out for Sn particles in a vacuum environment. The total area mass was \(20\; \rm {mg/cm^2}\). The velocity coefficient \(\beta = 10\; \rm{mg/cm^2}\) and the size coefficient \(\sigma = 0.5\).}
\label{fig9}
\end{figure}

\begin{figure}
\centering
\includegraphics[height=0.35\textwidth]{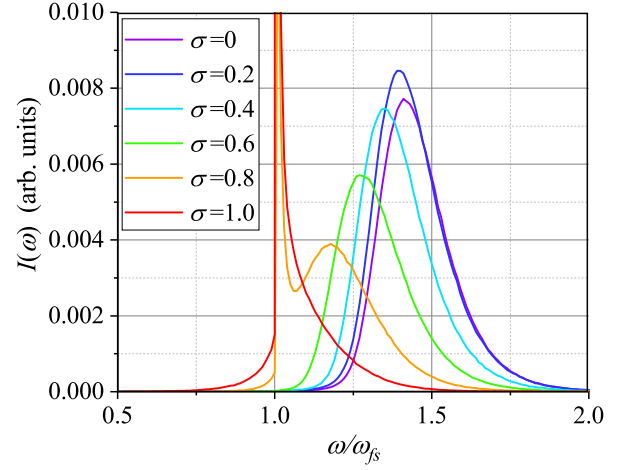}
\caption{Simulated PDV spectra with different particle size coefficients \(\sigma\). The calculation was carried out for Sn particles in a vacuum environment. The total area mass was \(20\; \rm {mg/cm^2}\). The velocity coefficient \(\beta = 10\) and the median diameter \(d_m = 1.5\; \rm \mu m\).}
\label{fig10}
\end{figure}

The results of numerical calculations that demonstrate the sensitivity of the PDV spectrum to changes in the ejecta parameters (\(\beta\), \(m_0\), \(d_m\), and \(\sigma\)) are presented in Figs.~\ref{fig7}--\ref{fig10}. The PDV spectra were simulated using the MC algorithm described in the previous section for Sn particles in a vacuum environment. The initial ejecta parameters were set to \(\beta=10\), \(m_0=20\; \rm{mg/cm^2}\), \(d_m=1.5\; \rm{\mu m}\), and \(\sigma=0.5\). In each figure, one of the parameters changes and the others remain constant.

The simulated PDV spectra with different values of the velocity coefficient \(\beta\) are shown in Fig.~\ref{fig7}. With an increase in \(\beta\), the spectrum peak moves towards the low velocities and its magnitude decreases. Furthermore, the spectrum shape becomes sharper and the high-velocity part of the spectrum becomes invisible. The coefficient \(\beta\) determines the distribution of particles in the ejecta. With larger \(\beta\), fewer particles are located at the top of the ejecta and the incident light can penetrate deeper. This results in the movement of the spectrum peak and a decrease in the observability of high-velocity particles.

The simulated PDV spectra with different values of the total area mass \(m_0\) are shown in Fig.~\ref{fig8}. The changes in the spectra can be divided into two sections. When \(m_0 \ge 10\; \rm mg/cm^2\), the spectrum displays a single peak. With a decrease in \(m_0\), this peak moves to the left, and its magnitude and slope exhibit slight changes. When \(m_0 \le 5\; \rm mg/cm^2\), a new peak appears around the free surface, and the spectrum exhibits a double-peak shape. A smaller area mass produces a more remarkable new peak. For \(m_0=2\; \rm mg/cm^2\), the original peak disappears and the spectrum again exhibits a single peak. The double-peak spectrum has been observed in previous experiments\cite{La:1,Andriyash:2} and simulations,\cite{Arsenii:1,Franzkowiak:2} and is the result of the direct exposure of incident light at the free surface.

There are two parameters that determine the particle size distribution---the median diameter \(d_m\) and the distribution width \(\sigma\). Their influence on the PDV spectrum is illustrated in Figs.~\ref{fig9} and \ref{fig10}, respectively. \(d_m\) and \(\sigma\) exhibit similar effects: as \(d_m\) or \(\sigma\) increases, the original peak of the spectrum moves towards the low velocities and the peak value decreases. A new peak then appears in the position of the free surface and the original peak gradually attenuates. This change in the form of the spectrum peak is similar to that for the area mass.

\subsection{Theoretical optical model}
The simulations described above using the MC algorithm provide a qualitative understanding of the influence of the ejecta parameters on the PDV spectrum. However, how to solve the reverse problem, i.e., extracting the ejecta parameters from the PDV spectrum, remains unclear. To obtain the quantitative relationships between the ejecta parameters and the characteristics of the PDV spectrum, we introduce the single-scattering theory. In this theory, the light is assumed to be scattered only once, and the scattering direction is always backward. With this assumption, the light direction is always parallel to the motion of the particles. The frequency shift of the light is proportional to the particle velocity, $\omega=2\omega_0 \cdot v/c$. The PDV spectrum can be expressed in terms of velocity, $I (\tilde v)$.

The above calculations have shown that the single scattering leads to an overestimation of the optical thickness. Here, we assume that the extinction process of particles only includes the backscattering and absorption effects, and the forward scattering is ignored. With this assumption, the PDV spectrum is calculated by the formula:

\begin{equation}
\begin{split}
I\left( \omega \right) &= I\left( {\tilde v} \right) \propto \sum\limits_i^{\tilde v < v < \tilde v + dv} {\sqrt {{I_{back,i}}} } \\
&= \sum\limits_i^{\tilde v < v < \tilde v + dv} {\sqrt {\frac{{{I_0}}}{{{A_s}}}{A_{back,i}}\left( v \right) \cdot \exp \left( { - 2\int_{\tilde v}^\infty {\tau \left( v \right)dv} } \right)} } \\
&\propto \sum\limits_i^{\tilde v < v < \tilde v + dv} {\sqrt {{A_{back,i}}\left( v \right)} } \cdot \sqrt {\exp \left( { - 2\int_{\tilde v}^\infty {\tau \left( v \right)dv} } \right)}
\end{split}
\label{eq22}
\end{equation}
where \(\tilde v\) is the velocity corresponding to the frequency shift, \(\tilde v=2\omega c/ \omega_r\), and \(A_{back}\) is the backscattering cross-section area of particles with velocity \(\tilde v\):

\begin{equation}
\begin{split}
&\sum\limits_i^{\tilde v < v < \tilde v + dv} {\sqrt {{A_{back,i}}\left( v \right)} } = \sum\limits_i^{\tilde v < v < \tilde v + dv} {\sqrt {\frac{1}{4}\pi d_i^2{K_{back}}} } \\
&= N\left( {\tilde v} \right)\frac{1}{2}\bar d\sqrt {\pi {K_{back}}} = \frac{{m\left( {\tilde v} \right){A_s}}}{{\frac{1}{6}\pi \overline {{d^3}} {\rho _0}}}\frac{1}{2}\bar d\sqrt {\pi {K_{back}}} \\
&= \frac{{3{A_s}\sqrt {{K_{back}}} }}{{\sqrt \pi {\rho _0}}}\frac{{{m_0}}}{{\overline {{d^3}} /\overline d }}f(\tilde v)
\end{split}
\label{eq23}
\end{equation}
where \(\overline {d}\) is the average particle diameter and \(K_{back}\) is the backscattering coefficient, which is calculated by Mie theory.

The exponent in Eq.~(22) denotes the extinction effects, where the coefficient 2 signifies the back and forth of light in the ejection process. The integral represents the contributions from the extinction of particles above the layer of velocity \(\tilde v\):

\begin{equation}
\begin{split}
\int_{\tilde v}^\infty {\tau \left( v \right)} dv &= \frac{{\sum\limits_i^{v > \tilde v} {\frac{\pi }{4}d_i^2{K^*_{ext}}} }}{{{A_s}}} = \frac{{\int_{\tilde v}^\infty {N\left( v \right)} dv \cdot \frac{\pi }{4}\overline {{d^2}} {K^*_{ext}}}}{{{A_s}}}\\
&= \frac{{3{K^*_{ext}}}}{{2{\rho _0}}}\frac{{\int_{\tilde v}^\infty {m\left( v \right)} dv}}{{\overline {{d^3}} /\overline {{d^2}} }}
\end{split}
\label{eq24}
\end{equation}
where the equivalent extinction coefficient \(K^*_{ext}\) only considers the backscattering and absorption effects:
\begin{equation}
K^*_{ext}=K_{back}+K_{abs}= g_{back}K_{ext}
\label{eq25}
\end{equation}
where the coefficient \(g_{back}\) is 0.5--0.7 for particle diameters of 1--10 \( \rm \mu m\). When \(g_{back}\) = 1, the present model reduces to that of Franzkowiak et al.

Because the velocity profile is exponential, \({\int_{\tilde v}^\infty {m\left( v \right)} dv} = m_0 f(\tilde v) v_{fs}/\beta\). Equation (\ref{eq25}) has the form:

\begin{equation}
\int_{\tilde v}^\infty {\tau \left( v \right)} dv = \frac{{3{K^*_{ext}}}}{{2{\rho _0}}}\frac{{{m_0}}}{{\overline {{d^3}} /\overline {{d^2}} }}\frac{v_{fs} }{{{\beta}}}f(v)
\label{eq26}
\end{equation}

When \(\tilde v\) is equal to the free surface velocity, the integral represents the amount of light that is able to reach the free surface. The optical thickness of the ejecta is defined as:

\begin{equation}
{\tau _0} = \int_{{v_{fs}}}^\infty {\tau \left( v \right)} dv = \frac{{3{K^*_{ext}}}}{{2{\rho _0}}}\frac{{{m_0}}}{{\overline {{d^3}} /\overline {{d^2}} }}
\label{eq27}
\end{equation}

Combining Eqs.~(\ref{eq23}), (\ref{eq26}), and (\ref{eq27}), we can write Eq.~(\ref{eq22}) as:
\begin{equation}
I\left( {\tilde v} \right) \propto \frac{{\sqrt {{K_{back}}} }}{{K_{ext}^*}} \cdot \frac{{\overline d}}{{\overline {{d^2}} }} \cdot {\tau _0}f\left( {\tilde v} \right)\exp \left( { - \frac{{{v_{fs}}}}{\beta }{\tau _0}f\left( {\tilde v} \right)} \right)
\label{eq28}
\end{equation}

\begin{figure}
\centering
\includegraphics[height=0.35\textwidth]{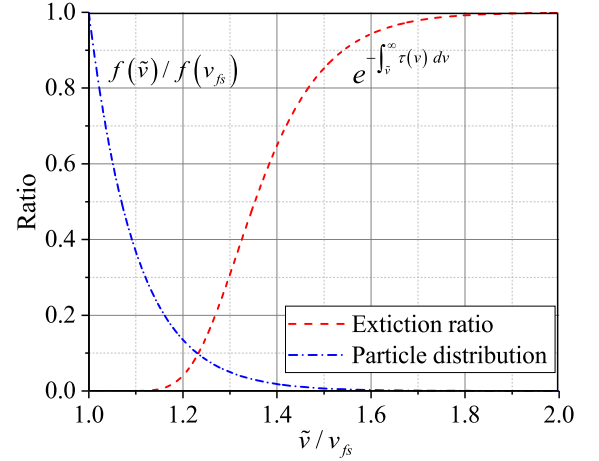}
\caption{Extinction ratio and particle distribution with respect to velocity.}
\label{fig11}
\end{figure}

This equation provides the theoretical form for determining the PDV spectrum from the ejecta parameters in a vacuum environment. In this formula, the PDV spectrum is proportional to the product of the velocity profile and the extinction term. These two parts are illustrated in Fig.~\ref{fig11}. As the velocity decreases, the corresponding ejecta position moves closer to the free surface, and the incident light becomes weaker because of particle extinction. However, the particles become dense deeper within the ejecta, and this enlarges the cross-section area of scattering. With the contribution of these two parts, the PDV spectrum exhibits a single peak shape.

\begin{figure}
\centering
\includegraphics[height=0.35\textwidth]{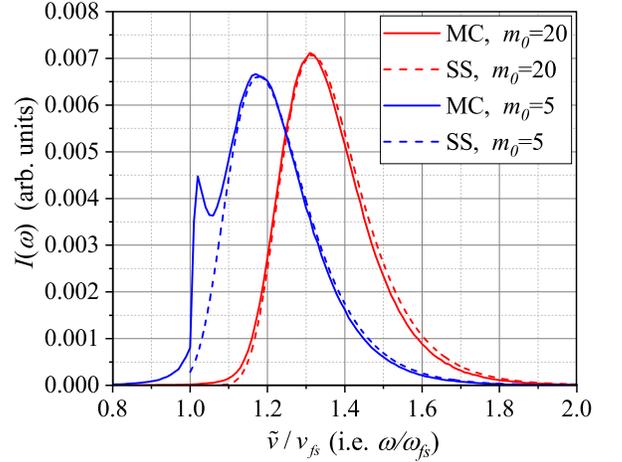}
\caption{Simulated PDV spectrum by MC algorithm and single-scattering (SS) model. The area mass unit is \( \rm mg/cm^2\). The particle settings are \(\beta=10\), \(d_m=1.5\; \rm \mu m\), \(\sigma=0.5\). The backscattering coefficient is \(g_{back}=0.67\). }
\label{fig12}
\end{figure}

The simulated PDV spectra of the present model are compared with those of the MC algorithm in Fig.~\ref{fig12}. With the correction of the backscattering coefficient, there is a good agreement between the results, both in the main peak position and the curve shape. However, in the case of a small ejecta mass (\(m_0=5\; \rm mg/cm^2\)), the present model cannot simulate the peak around the free surface. Although the spectrum is nonzero at the position of the free surface in this model, neglecting the multiple-scattering results prevents the second peak from appearing.\cite{Arsenii:1,Franzkowiak:2} In this paper, we mainly consider the information supplied by the original peak of the PDV spectrum. Thus, this defect has only a very slight influence on the accuracy of the present model.

We now analyze the spectrum function [Eq.~(\ref{eq28})]. First, we take its derivative:
\begin{equation}
\begin{split}
I'\left( {\tilde v} \right) &\propto {\tau _0}f'\left( {\tilde v} \right)\exp \left( { - \frac{{{v_{fs}}}}{\beta }{\tau _0}f\left( {\tilde v} \right)} \right)\left( {1 - \frac{{{v_{fs}}}}{\beta }{\tau _0}f\left( {\tilde v} \right)} \right)\\
&= - \frac{\beta }{{{v_{fs}}}}I\left( {\tilde v} \right)\left( {1 - \frac{{{v_{fs}}}}{\beta }{\tau _0}f\left( {\tilde v} \right)} \right)
\end{split}
\label{eq29}
\end{equation}
where \(f'(\tilde v)=-\beta/v_{fs}f(\tilde v)\).

When \(I'(\tilde v) = 0\), the solution provides the position of the spectrum peak:
\begin{equation}
1 - \frac{{{v_{fs}}}}{\beta }{\tau _0}f\left( {{{\tilde v}_{peak}}} \right) = 0
\label{eq30}
\end{equation}
\begin{equation}
{{\tilde v}_{peak}}/{v_{fs}} = 1 + \ln \left( {{\tau _0}} \right)/\beta
\label{eq31}
\end{equation}

The peak value of the PDV spectrum is:
\begin{equation}
I\left( {{{\tilde v}_{peak}}} \right) \propto \frac{{\sqrt {{K_{back}}} }}{{K_{ext}^*}} \cdot \frac{{\bar d}}{{\overline {{d^2}} }} \cdot \frac{\beta }{{{v_{fs}}}}{e^{ - 1}} \propto \frac{{\beta}}{{\overline {{d^2}} / \overline {{d}} }}
\label{eq32}
\end{equation}

Finally, the relative curvature at the spectrum peak is given by the second derivative of the spectrum function:
\begin{equation}
\begin{split}
I''\left( {\tilde v} \right)/I\left( {{{\tilde v}_{peak}}} \right) = \frac{{{v_{fs}}}}{\beta }e{\left( {{\tau _0}f\left( {\tilde v} \right)\exp \left( { - \frac{{{v_{fs}}}}{\beta }{\tau _0}f\left( {\tilde v} \right)} \right)} \right)^{\prime \prime }}
\end{split}
\label{eq33}
\end{equation}

\begin{equation}
I''\left( {{{\tilde v}_{peak}}} \right)/I\left( {{{\tilde v}_{peak}}} \right) = - {\left( {\frac{\beta }{{{v_{fs}}}}} \right)^2}
\label{eq34}
\end{equation}

In summary, the relationships between the ejecta parameters and the characteristics of the PDV spectrum have the form:
\begin{equation}
\left\{
\begin{aligned}
&{{\tilde v}_{peak}}/{v_{fs}} = 1 + \ln \left( {{\tau _0}} \right)/\beta \\
&I\left( {{{\tilde v}_{peak}}/{v_{fs}}} \right) \propto \frac{\beta }{{\overline {{d^2}} /\overline d }}\\
&I''\left( {{{\tilde v}_{peak}}/{v_{fs}}} \right)/I\left( {{{\tilde v}_{peak}}/{v_{fs}}} \right) = - {\beta ^2}
\end{aligned}
\right.
\label{eq35}
\end{equation}
where the independent variable is normalized by the velocity of the free surface, \(v_{fs}\).

The above relationships allow some of the ejecta parameters to be determined. First, the velocity profile coefficient \(\beta\) is given by the relative curvature at the spectrum peak. In addition, the surface mean diameter \(d^*=\overline d^2/ \overline {d}=d_m e^{0.5\sigma^2}\) can be derived from the value of the spectrum peak. Finally, the optical thickness of the ejecta \(\tau_0\) is obtained from the position of the spectrum peak, where \(\tau_0\) is related to the ejecta area mass \(m_0\) and the Sauter mean diameter \(d_s=\overline {d^3}/ \overline {d^2}=d_m e^{2.5\sigma^2}\). If \(d_s\) is assumed to be approximately \(d^*\), the area mass \(m_0\) can be determined.

Figures 13--15 compare these theoretical relationships with the MC simulation results. Two optical thickness and a large range of velocity profiles are considered. It can be observed that the theoretical relationships largely conform to the MC simulations in these cases, which verifies the present model to some extent.

\begin{figure}
\centering
\includegraphics[height=0.35\textwidth]{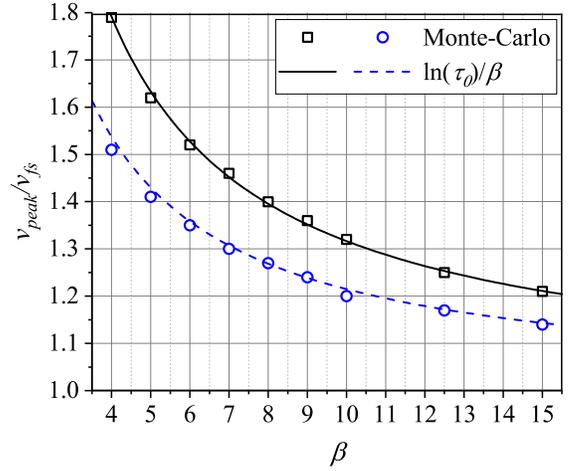}
\caption{Peak positions of PDV spectrum calculated by MC algorithm and theoretical formula. Two optical thickness are considered, where the square points denote \(\tau_0=23.4\) and the circular points denote \(\tau_0=8.6\). The corresponding ejecta parameters are $m_0 =20\;{\rm mg/cm^2}, d_m=1.5\; \mu{\rm m}, \sigma = 0.5$ and $m_0 =10\;{\rm mg/cm^2}, d_m=2.0\; \mu{\rm m}, \sigma = 0.5$, respectively. }
\label{fig13}
\end{figure}

\begin{figure}
\centering
\includegraphics[height=0.35\textwidth]{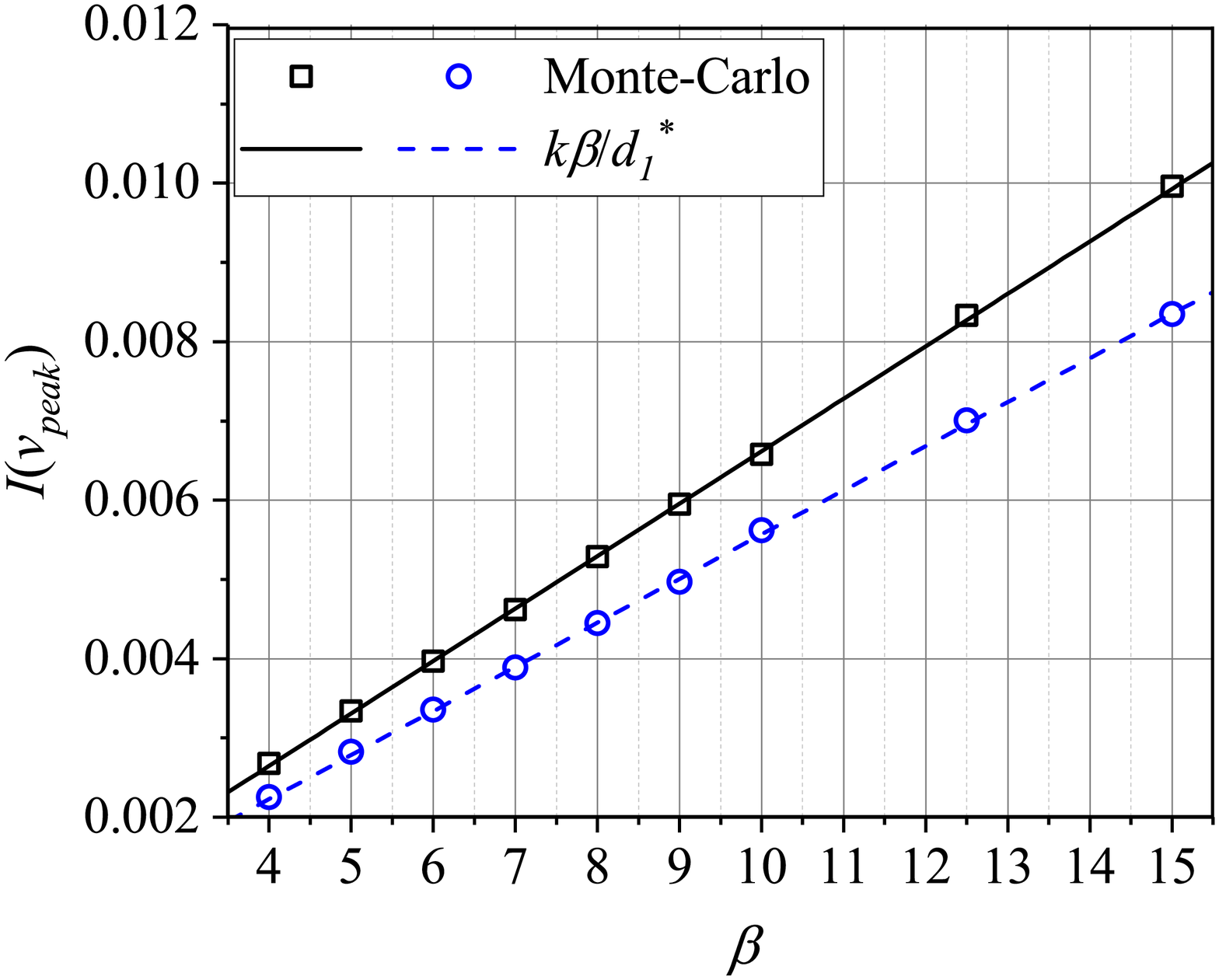}
\caption{Peak values of PDV spectrum calculated by MC algorithm and theoretical formula. Two optical thickness are considered, where the square points denote \(\tau_0=23.4\) and the circular points denote \(\tau_0=8.6\). The corresponding ejecta parameters are $m_0 =20\;{\rm mg/cm^2}, d_m=1.5\; \mu{\rm m}, \sigma = 0.5$ and $m_0 =10\;{\rm mg/cm^2}, d_m=2.0\; \mu{\rm m}, \sigma = 0.5$, respectively. }
\label{fig14}
\end{figure}

\begin{figure}
\centering
\includegraphics[height=0.35\textwidth]{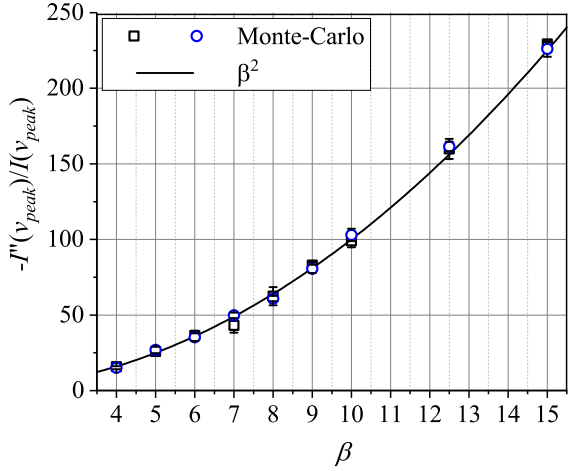}
\caption{Relative curvature of PDV spectrum at the peak calculated by MC algorithm and theoretical formula. Two optical thickness are considered, where the square points denote \(\tau_0=23.4\) and the circular points denote \(\tau_0=8.6\). The corresponding ejecta parameters are $m_0 =20\;{\rm mg/cm^2}, d_m=1.5\; \mu{\rm m}, \sigma = 0.5$ and $m_0 =10\;{\rm mg/cm^2}, d_m=2.0\; \mu{\rm m}, \sigma = 0.5$, respectively.}
\label{fig15}
\end{figure}

\subsection{Experimental verification}
In the above relationships, the peak value of the spectrum is difficult to use in the analysis of PDV experiments. In the experiments, the PDV spectrum is scaled by the reference light intensity, probe reception, photoelectric conversion efficiency, and circuit amplification factor, among other factors. Additional PDV experiments are required to calibrate this scaled factor. For a single PDV vacuum experiment, only the velocity profile \(\beta\) and optical thickness \(\tau_0\) of the ejecta can be extracted. If there is an additional particle granularity measurement, the ejecta area mass \(m_0\) can also be determined.

\begin{figure}
\centering
\includegraphics[height=0.40\textwidth]{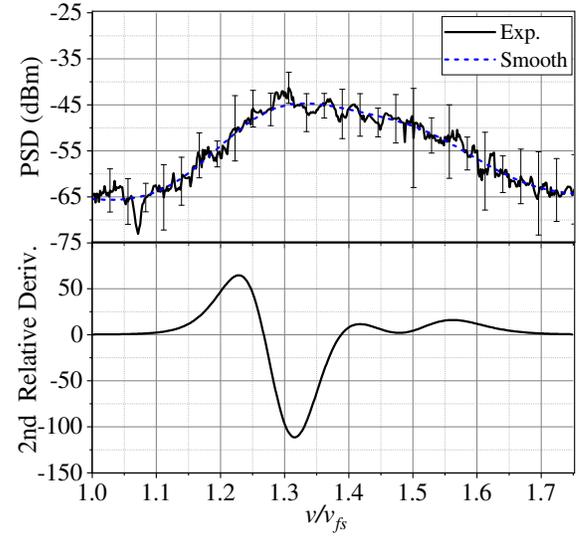}
\caption{(a) PDV spectrogram extracted from ejecta experiment of Franzkowiak et al.~\cite{Franzkowiak:1} and (b) second derivative of the smooth data.}
\label{fig16}
\end{figure}

A set of ejecta PDV experiments performed by Franzkowiak et al.\cite{Franzkowiak:1} was used to verify the present theoretical model. The experiment was carried out in a vacuum environment using Sn material with the surface machined into \(60 \times 8\; \rm \mu m\) grooves. The shock-induced breakout pressure was \(P_{SB}=28\; \rm GPa\). The velocity of the free surface was found to be approximately \(2013\; \rm m/s\). We extracted the PDV spectrum from the experimental spectrogram over the period \(0.2-0.8\; \rm \mu s\), as shown in Fig.~\ref{fig16}(a). The PDV data were then averaged and smoothed using the low-pass filtering of the fast Fourier transform. We converted the spectrum units [dBm] to volts and then took the second derivative to give the smoothed PDV spectrum shown in Fig.~\ref{fig16}(b). The peak of this spectrum is located at \(\tilde v /v_{fs} = 1.32\) and the corresponding relative curvature is approximately $-110$. Combined with Eq.~(\ref{eq35}), this suggests a velocity profile coefficient of \(\beta=10.5\) and an optical thickness of \(\tau_0=28.79\).

Schauer et al.\cite{Schauer:1} conducted a Mie-scattering experiment with similar conditions, where the surface roughness was \(50 \times 8\; \mu m\) and the breakout pressure was about \(30\; \rm GPa\). The particle size distribution was measured to be \(d_m=0.6\;\mu{\rm{m}}, \sigma=0.5\). Using this data, the total area mass was determined to be \(m_0=7.5\;{\rm{mg/cm^2}}\).

\begin{figure}
\centering
\includegraphics[height=0.30\textwidth]{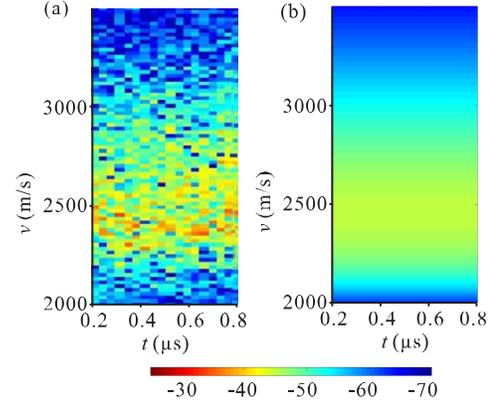}
\caption{Comparison of the experimental and simulated PDV spectra.}
\label{fig17}
\end{figure}

\begin{figure}
\centering
\includegraphics[height=0.35\textwidth]{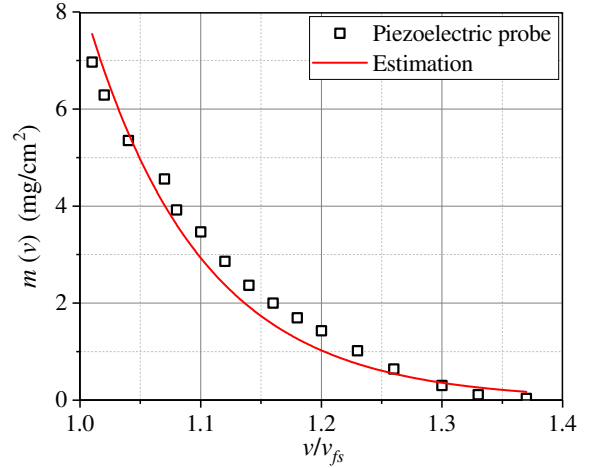}
\caption{Comparison of the area mass and velocity profile between the piezoelectric probe measurement~\cite{Franzkowiak:1} and our estimation.}
\label{fig18}
\end{figure}

In their PDV experiment, Franzkowiak et al. simultaneously measured the area mass with respect to velocity using a piezoelectric probe. The PDV spectrum and area mass given by our estimations and their experiments are compared in Figs.~\ref{fig17} and~\ref{fig18}, respectively. These two results are in good agreement, which verifies the present theoretical model.

\section{Conclusion}
This paper has discussed the PDV spectrum of ejecta particles from shock-loaded samples in a vacuum. A GPU-accelerated MC algorithm that rebuilds the PDV spectrum for the ejecta particles has been proposed, and Mie theory was applied to describe the scattering process. Compared with the reconstruction methods of Andriyash et al. and Franzkowiak et al., a reasonable scattering model is the key to simulating the PDV spectrum accurately. The simulations using the MC algorithm indicate that the particle velocity profile, particle size, and ejecta area mass have a significant influence on the shape and values of the PDV spectrum. As the velocity profile coefficient or particle size increases, the spectrum peak moves to lower velocities. However, this change in the spectrum peak is reversed for the total area mass. In addition, for small values of the optical thickness (few ejecta mass or large particle size), a new spectrum peak appears near the free surface and the original peak gradually decreases or even disappears. For a quantitative analysis, a corrected single-scattering model was proposed for deriving the relationships between the ejecta parameters and the characteristics of the PDV spectrum. It was found that the relative curvature of the spectrum peak is equal to the square of the velocity profile coefficient \(\beta\). The peak value of the spectrum is proportional to the ratio of \(\beta\) to the particle size \(d\), and the peak position of the spectrum is related to \(\beta\) and the total extinction coefficient \(\tau_0\), where \(\tau_0\) is calculated from the total area mass \(m_0\) and the particle size \(d\). Thus, the ejecta parameters \(\beta\), \(d\), and \(m_0\) can be resolved using information about the spectrum peak. However, the spectrum is scaled by multiple experimental parameters, and the relationship with the particle size is difficult to determine. For a single PDV spectrum, only the velocity profile and optical thickness can be determined. Finally, the theoretical interpretation was found to be in good agreement with the MC simulations and PDV experiments of ejecta in a vacuum environment.

The present theoretical model does not consider the multiple scattering near the free surface. When the optical thickness is sufficiently small and the original peak disappears, the present model may be invalid. How to determine the particle size in the PDV experiment is another unsolved issue. In a gas environment, the particles slow down because of aerodynamic deceleration, and this introduces a series of changes to the PDV spectrum over time. The particle deceleration is related to the particle size. In future work, the PDV spectrum in a gas environment will be discussed in an attempt to recover more comprehensive quantities of the ejecta.

\begin{acknowledgments}
This work was supported by a joint fund from the National Natural Science Foundation of China (Grant Nos. 11902043, 11772065) and the Science Challenge Project (Grant No. TZ2016001).
\end{acknowledgments}

\section*{Data Availability}
The data that support the findings of this study are available from the corresponding author
upon reasonable request.

\nocite{*}
\bibliography{aipsamp}%

\begin{thebibliography}{42}%
\makeatletter
\providecommand \@ifxundefined [1]{%
 \@ifx{#1\undefined}
}%
\providecommand \@ifnum [1]{%
 \ifnum #1\expandafter \@firstoftwo
 \else \expandafter \@secondoftwo
 \fi
}%
\providecommand \@ifx [1]{%
 \ifx #1\expandafter \@firstoftwo
 \else \expandafter \@secondoftwo
 \fi
}%
\providecommand \natexlab [1]{#1}%
\providecommand \enquote  [1]{``#1''}%
\providecommand \bibnamefont  [1]{#1}%
\providecommand \bibfnamefont [1]{#1}%
\providecommand \citenamefont [1]{#1}%
\providecommand \href@noop [0]{\@secondoftwo}%
\providecommand \href [0]{\begingroup \@sanitize@url \@href}%
\providecommand \@href[1]{\@@startlink{#1}\@@href}%
\providecommand \@@href[1]{\endgroup#1\@@endlink}%
\providecommand \@sanitize@url [0]{\catcode `\\12\catcode `\$12\catcode
  `\&12\catcode `\#12\catcode `\^12\catcode `\_12\catcode `\%12\relax}%
\providecommand \@@startlink[1]{}%
\providecommand \@@endlink[0]{}%
\providecommand \url  [0]{\begingroup\@sanitize@url \@url }%
\providecommand \@url [1]{\endgroup\@href {#1}{\urlprefix }}%
\providecommand \urlprefix  [0]{URL }%
\providecommand \Eprint [0]{\href }%
\providecommand \doibase [0]{http://dx.doi.org/}%
\providecommand \selectlanguage [0]{\@gobble}%
\providecommand \bibinfo  [0]{\@secondoftwo}%
\providecommand \bibfield  [0]{\@secondoftwo}%
\providecommand \translation [1]{[#1]}%
\providecommand \BibitemOpen [0]{}%
\providecommand \bibitemStop [0]{}%
\providecommand \bibitemNoStop [0]{.\EOS\space}%
\providecommand \EOS [0]{\spacefactor3000\relax}%
\providecommand \BibitemShut  [1]{\csname bibitem#1\endcsname}%
\let\auto@bib@innerbib\@empty
\bibitem [{\citenamefont {Sollier}\ and\ \citenamefont
  {Lescoute}(2020)}]{Sollier:1}%
  \BibitemOpen
  \bibfield  {author} {\bibinfo {author} {\bibfnamefont {A.}~\bibnamefont
  {Sollier}}\ and\ \bibinfo {author} {\bibfnamefont {E.}~\bibnamefont
  {Lescoute}},\ }\bibfield  {title} {\enquote {\bibinfo {title}
  {Characterization of the ballistic properties of ejecta from laser
  shock-loaded samples using high resolution picosecond laser imaging},}\
  }\href@noop {} {\bibfield  {journal} {\bibinfo  {journal} {Int. J. Impact
  Eng.}\ }\textbf {\bibinfo {volume} {136}},\ \bibinfo {pages} {103429}
  (\bibinfo {year} {2020})}\BibitemShut {NoStop}%
\bibitem [{\citenamefont {Monfared}\ \emph {et~al.}(2014)\citenamefont
  {Monfared}, \citenamefont {Oro}, \citenamefont {Graver}, \citenamefont
  {Hammerberg}, \citenamefont {Lalone}, \citenamefont {Pack}, \citenamefont
  {Schauer}, \citenamefont {Stevens}, \citenamefont {Stone},\ and\
  \citenamefont {Turley}}]{Monfared:1}%
  \BibitemOpen
  \bibfield  {author} {\bibinfo {author} {\bibfnamefont {S.~K.}\ \bibnamefont
  {Monfared}}, \bibinfo {author} {\bibfnamefont {D.~M.}\ \bibnamefont {Oro}},
  \bibinfo {author} {\bibfnamefont {M.}~\bibnamefont {Graver}}, \bibinfo
  {author} {\bibfnamefont {J.~E.}\ \bibnamefont {Hammerberg}}, \bibinfo
  {author} {\bibfnamefont {B.~M.}\ \bibnamefont {Lalone}}, \bibinfo {author}
  {\bibfnamefont {C.~L.}\ \bibnamefont {Pack}}, \bibinfo {author}
  {\bibfnamefont {M.~M.}\ \bibnamefont {Schauer}}, \bibinfo {author}
  {\bibfnamefont {G.~D.}\ \bibnamefont {Stevens}}, \bibinfo {author}
  {\bibfnamefont {J.~B.}\ \bibnamefont {Stone}}, \ and\ \bibinfo {author}
  {\bibfnamefont {W.~D.~a.}\ \bibnamefont {Turley}},\ }\bibfield  {title}
  {\enquote {\bibinfo {title} {Experimental observations on the links between
  surface perturbation parameters and shock-induced mass ejection},}\
  }\href@noop {} {\bibfield  {journal} {\bibinfo  {journal} {J. Appl. Phys.}\
  }\textbf {\bibinfo {volume} {116}},\ \bibinfo {pages} {063504} (\bibinfo
  {year} {2014})}\BibitemShut {NoStop}%
\bibitem [{\citenamefont {Asay}, \citenamefont {Mix},\ and\ \citenamefont
  {Perry}(1976)}]{Asay:1}%
  \BibitemOpen
  \bibfield  {author} {\bibinfo {author} {\bibfnamefont {J.~R.}\ \bibnamefont
  {Asay}}, \bibinfo {author} {\bibfnamefont {L.~P.}\ \bibnamefont {Mix}}, \
  and\ \bibinfo {author} {\bibfnamefont {F.~C.}\ \bibnamefont {Perry}},\
  }\bibfield  {title} {\enquote {\bibinfo {title} {Ejection of material from
  shocked surfaces},}\ }\href {\doibase 10.1063/1.89066} {\bibfield  {journal}
  {\bibinfo  {journal} {Appl. Phys. Lett.}\ }\textbf {\bibinfo {volume} {29}},\
  \bibinfo {pages} {284} (\bibinfo {year} {1976})}\BibitemShut {NoStop}%
\bibitem [{\citenamefont {Speight}, \citenamefont {Harper},\ and\ \citenamefont
  {Smeeton}(1989)}]{Speight:1}%
  \BibitemOpen
  \bibfield  {author} {\bibinfo {author} {\bibfnamefont {C.~S.}\ \bibnamefont
  {Speight}}, \bibinfo {author} {\bibfnamefont {L.}~\bibnamefont {Harper}}, \
  and\ \bibinfo {author} {\bibfnamefont {V.~S.}\ \bibnamefont {Smeeton}},\
  }\bibfield  {title} {\enquote {\bibinfo {title} {Piezoelectric probe for the
  detection of shock-induced spray and spall},}\ }\href {\doibase
  10.1063/1.1140443} {\bibfield  {journal} {\bibinfo  {journal} {Rev. Sci.
  Instrum.}\ }\textbf {\bibinfo {volume} {60}},\ \bibinfo {pages} {3802}
  (\bibinfo {year} {1989})}\BibitemShut {NoStop}%
\bibitem [{\citenamefont {Ogorodnikov}\ \emph {et~al.}(1998)\citenamefont
  {Ogorodnikov}, \citenamefont {Ivanov}, \citenamefont {Mikhailov},
  \citenamefont {Kryukov},\ and\ \citenamefont {Golubev}}]{Ogorodnikov:1}%
  \BibitemOpen
  \bibfield  {author} {\bibinfo {author} {\bibfnamefont {V.~A.}\ \bibnamefont
  {Ogorodnikov}}, \bibinfo {author} {\bibfnamefont {A.~G.}\ \bibnamefont
  {Ivanov}}, \bibinfo {author} {\bibfnamefont {A.~L.}\ \bibnamefont
  {Mikhailov}}, \bibinfo {author} {\bibfnamefont {N.~I.}\ \bibnamefont
  {Kryukov}}, \ and\ \bibinfo {author} {\bibfnamefont {V.~A.}\ \bibnamefont
  {Golubev}},\ }\bibfield  {title} {\enquote {\bibinfo {title} {Particle
  ejection from the shocked free surface of metals and diagnostic methods for
  these particles},}\ }\href {\doibase 10.1007/BF02672705} {\bibfield
  {journal} {\bibinfo  {journal} {Combust. Explos. Shock Waves}\ }\textbf
  {\bibinfo {volume} {34}},\ \bibinfo {pages} {696} (\bibinfo {year}
  {1998})}\BibitemShut {NoStop}%
\bibitem [{\citenamefont {Bistow}\ and\ \citenamefont {Hyde}(1969)}]{Bistow:1}%
  \BibitemOpen
  \bibfield  {author} {\bibinfo {author} {\bibfnamefont {W.~F.}\ \bibnamefont
  {Bistow}}\ and\ \bibinfo {author} {\bibfnamefont {E.~F.}\ \bibnamefont
  {Hyde}},\ }\bibfield  {title} {\enquote {\bibinfo {title} {Surface spray from
  explosively accelerated metal plates as an indicator of melting},}\ }in\
  \href@noop {} {\emph {\bibinfo {booktitle} {The U.K. National Archives}}},\
  \bibinfo {series and number} {\bibinfo {number} {Technical Report ES
  4/1152}}\ (\bibinfo {year} {1969})\BibitemShut {NoStop}%
\bibitem [{\citenamefont {Richtmyer}(1960)}]{Richtmyer:1}%
  \BibitemOpen
  \bibfield  {author} {\bibinfo {author} {\bibfnamefont {R.~D.}\ \bibnamefont
  {Richtmyer}},\ }\bibfield  {title} {\enquote {\bibinfo {title} {Taylor
  instability in shock acceleration of compressible fluids},}\ }\href@noop {}
  {\bibfield  {journal} {\bibinfo  {journal} {Proc. Lond. Math. Soc.}\ }\textbf
  {\bibinfo {volume} {XIII}},\ \bibinfo {pages} {297--319} (\bibinfo {year}
  {1960})}\BibitemShut {NoStop}%
\bibitem [{\citenamefont {Meshkov}(1969)}]{Meshkov:1}%
  \BibitemOpen
  \bibfield  {author} {\bibinfo {author} {\bibfnamefont {E.~E.}\ \bibnamefont
  {Meshkov}},\ }\bibfield  {title} {\enquote {\bibinfo {title} {Instability in
  shock-accelerated boundary separating two gasses},}\ }\href@noop {}
  {\bibfield  {journal} {\bibinfo  {journal} {Izv. Akad. Nauk SSSR Mekh. Gaza}\
  }\textbf {\bibinfo {volume} {5}},\ \bibinfo {pages} {151--158} (\bibinfo
  {year} {1969})}\BibitemShut {NoStop}%
\bibitem [{\citenamefont {Yeager}\ \emph {et~al.}(2017)\citenamefont {Yeager},
  \citenamefont {Bowden}, \citenamefont {Guildenbecher},\ and\ \citenamefont
  {Olles}}]{Yeager:1}%
  \BibitemOpen
  \bibfield  {author} {\bibinfo {author} {\bibfnamefont {J.~D.}\ \bibnamefont
  {Yeager}}, \bibinfo {author} {\bibfnamefont {P.~R.}\ \bibnamefont {Bowden}},
  \bibinfo {author} {\bibfnamefont {D.~R.}\ \bibnamefont {Guildenbecher}}, \
  and\ \bibinfo {author} {\bibfnamefont {J.~D.}\ \bibnamefont {Olles}},\
  }\bibfield  {title} {\enquote {\bibinfo {title} {Characterization of
  hypervelocity metal fragments for explosive initiation},}\ }\href@noop {}
  {\bibfield  {journal} {\bibinfo  {journal} {J. Appl. Phys.}\ }\textbf
  {\bibinfo {volume} {122}},\ \bibinfo {pages} {035901} (\bibinfo {year}
  {2017})}\BibitemShut {NoStop}%
\bibitem [{\citenamefont {Held}(1996)}]{Held:1}%
  \BibitemOpen
  \bibfield  {author} {\bibinfo {author} {\bibfnamefont {M.}~\bibnamefont
  {Held}},\ }\bibfield  {title} {\enquote {\bibinfo {title} {Initiation
  criteria of high explosives at different projectile or jet densities},}\
  }\href@noop {} {\bibfield  {journal} {\bibinfo  {journal} {Propellants
  Explos. Pyrotech.}\ }\textbf {\bibinfo {volume} {21}},\ \bibinfo {pages}
  {235} (\bibinfo {year} {1996})}\BibitemShut {NoStop}%
\bibitem [{\citenamefont {Tokheim}\ \emph {et~al.}(1999)\citenamefont
  {Tokheim}, \citenamefont {Curran}, \citenamefont {Seaman}, \citenamefont
  {Cooper},\ and\ \citenamefont {Schirmann}}]{Tokheim:1}%
  \BibitemOpen
  \bibfield  {author} {\bibinfo {author} {\bibfnamefont {R.~E.}\ \bibnamefont
  {Tokheim}}, \bibinfo {author} {\bibfnamefont {D.~R.}\ \bibnamefont {Curran}},
  \bibinfo {author} {\bibfnamefont {L.}~\bibnamefont {Seaman}}, \bibinfo
  {author} {\bibfnamefont {T.}~\bibnamefont {Cooper}}, \ and\ \bibinfo {author}
  {\bibfnamefont {D.}~\bibnamefont {Schirmann}},\ }\bibfield  {title} {\enquote
  {\bibinfo {title} {Hypervelocity shrapnel damage assessment in the nif target
  chamber},}\ }\href@noop {} {\bibfield  {journal} {\bibinfo  {journal} {Int.
  J. Impact Eng.}\ }\textbf {\bibinfo {volume} {23}},\ \bibinfo {pages} {933}
  (\bibinfo {year} {1999})}\BibitemShut {NoStop}%
\bibitem [{\citenamefont {Masters}\ \emph {et~al.}(2016)\citenamefont
  {Masters}, \citenamefont {Fisher}, \citenamefont {Kalantar}, \citenamefont
  {St?lken}, \citenamefont {Smith}, \citenamefont {Vignes}, \citenamefont
  {Burns}, \citenamefont {Doeppner}, \citenamefont {Kritcher},\ and\
  \citenamefont {Park}}]{Masters:1}%
  \BibitemOpen
  \bibfield  {author} {\bibinfo {author} {\bibfnamefont {N.~D.}\ \bibnamefont
  {Masters}}, \bibinfo {author} {\bibfnamefont {A.}~\bibnamefont {Fisher}},
  \bibinfo {author} {\bibfnamefont {D.}~\bibnamefont {Kalantar}}, \bibinfo
  {author} {\bibfnamefont {J.}~\bibnamefont {St?lken}}, \bibinfo {author}
  {\bibfnamefont {C.}~\bibnamefont {Smith}}, \bibinfo {author} {\bibfnamefont
  {R.}~\bibnamefont {Vignes}}, \bibinfo {author} {\bibfnamefont
  {S.}~\bibnamefont {Burns}}, \bibinfo {author} {\bibfnamefont
  {T.}~\bibnamefont {Doeppner}}, \bibinfo {author} {\bibfnamefont
  {A.}~\bibnamefont {Kritcher}}, \ and\ \bibinfo {author} {\bibfnamefont
  {H.~S.}\ \bibnamefont {Park}},\ }\bibfield  {title} {\enquote {\bibinfo
  {title} {Debris and shrapnel assessments for {N}ational {I}gnition {F}acility
  targets and diagnostics},}\ }\href@noop {} {\bibfield  {journal} {\bibinfo
  {journal} {J. Phys.: Conf. Ser.}\ }\textbf {\bibinfo {volume} {717}},\
  \bibinfo {pages} {012108} (\bibinfo {year} {2016})}\BibitemShut {NoStop}%
\bibitem [{\citenamefont {Asay}(1978)}]{Asay:2}%
  \BibitemOpen
  \bibfield  {author} {\bibinfo {author} {\bibfnamefont {J.~R.}\ \bibnamefont
  {Asay}},\ }\bibfield  {title} {\enquote {\bibinfo {title} {Thick-plate
  technique for measuring ejecta from shocked surfaces},}\ }\href {\doibase
  10.1063/1.324545} {\bibfield  {journal} {\bibinfo  {journal} {J. Appl.
  Phys.}\ }\textbf {\bibinfo {volume} {49}},\ \bibinfo {pages} {6173} (\bibinfo
  {year} {1978})}\BibitemShut {NoStop}%
\bibitem [{\citenamefont {He}\ \emph {et~al.}(2014)\citenamefont {He},
  \citenamefont {Xin}, \citenamefont {Chu}, \citenamefont {Li},\ and\
  \citenamefont {Gu}}]{He:1}%
  \BibitemOpen
  \bibfield  {author} {\bibinfo {author} {\bibfnamefont {W.}~\bibnamefont
  {He}}, \bibinfo {author} {\bibfnamefont {J.}~\bibnamefont {Xin}}, \bibinfo
  {author} {\bibfnamefont {G.}~\bibnamefont {Chu}}, \bibinfo {author}
  {\bibfnamefont {J.}~\bibnamefont {Li}}, \ and\ \bibinfo {author}
  {\bibfnamefont {Y.}~\bibnamefont {Gu}},\ }\bibfield  {title} {\enquote
  {\bibinfo {title} {Investigation of fragment sizes in laser-driven
  shock-loaded tin with improved watershed segmentation method},}\ }\href@noop
  {} {\bibfield  {journal} {\bibinfo  {journal} {Opt. Express}\ }\textbf
  {\bibinfo {volume} {22}},\ \bibinfo {pages} {18924} (\bibinfo {year}
  {2014})}\BibitemShut {NoStop}%
\bibitem [{\citenamefont {Vogan}\ \emph {et~al.}(2005)\citenamefont {Vogan},
  \citenamefont {Anderson}, \citenamefont {Grover}, \citenamefont {Hammerberg},
  \citenamefont {King}, \citenamefont {Lamoreaux}, \citenamefont {Macrum},
  \citenamefont {Morley}, \citenamefont {Rigg},\ and\ \citenamefont
  {Stevens}}]{Vogan:1}%
  \BibitemOpen
  \bibfield  {author} {\bibinfo {author} {\bibfnamefont {W.~S.}\ \bibnamefont
  {Vogan}}, \bibinfo {author} {\bibfnamefont {W.~W.}\ \bibnamefont {Anderson}},
  \bibinfo {author} {\bibfnamefont {M.}~\bibnamefont {Grover}}, \bibinfo
  {author} {\bibfnamefont {J.~E.}\ \bibnamefont {Hammerberg}}, \bibinfo
  {author} {\bibfnamefont {N.~S.~P.}\ \bibnamefont {King}}, \bibinfo {author}
  {\bibfnamefont {S.~K.}\ \bibnamefont {Lamoreaux}}, \bibinfo {author}
  {\bibfnamefont {G.}~\bibnamefont {Macrum}}, \bibinfo {author} {\bibfnamefont
  {K.~B.}\ \bibnamefont {Morley}}, \bibinfo {author} {\bibfnamefont {P.~A.}\
  \bibnamefont {Rigg}}, \ and\ \bibinfo {author} {\bibfnamefont {G.~D.~a.}\
  \bibnamefont {Stevens}},\ }\bibfield  {title} {\enquote {\bibinfo {title}
  {Piezoelectric characterization of ejecta from shocked tin surfaces},}\
  }\href@noop {} {\bibfield  {journal} {\bibinfo  {journal} {J. Appl. Phys.}\
  }\textbf {\bibinfo {volume} {98}},\ \bibinfo {pages} {284} (\bibinfo {year}
  {2005})}\BibitemShut {NoStop}%
\bibitem [{\citenamefont {Sorenson}\ \emph {et~al.}(2002)\citenamefont
  {Sorenson}, \citenamefont {Minich}, \citenamefont {Romero}, \citenamefont
  {Tunnell},\ and\ \citenamefont {Malone}}]{Sorenson:1}%
  \BibitemOpen
  \bibfield  {author} {\bibinfo {author} {\bibfnamefont {D.~S.}\ \bibnamefont
  {Sorenson}}, \bibinfo {author} {\bibfnamefont {R.~W.}\ \bibnamefont
  {Minich}}, \bibinfo {author} {\bibfnamefont {J.~L.}\ \bibnamefont {Romero}},
  \bibinfo {author} {\bibfnamefont {T.~W.}\ \bibnamefont {Tunnell}}, \ and\
  \bibinfo {author} {\bibfnamefont {R.~M.}\ \bibnamefont {Malone}},\ }\bibfield
   {title} {\enquote {\bibinfo {title} {Ejecta particle size distributions for
  shock loaded sn and al metals},}\ }\href@noop {} {\bibfield  {journal}
  {\bibinfo  {journal} {J. Appl. Phys.}\ }\textbf {\bibinfo {volume} {92}},\
  \bibinfo {pages} {5830--5836} (\bibinfo {year} {2002})}\BibitemShut {NoStop}%
\bibitem [{\citenamefont {Sorenson}\ \emph {et~al.}(2017)\citenamefont
  {Sorenson}, \citenamefont {Capelle}, \citenamefont {Grover}, \citenamefont
  {Johnson},\ and\ \citenamefont {Turley}}]{Sorenson:2}%
  \BibitemOpen
  \bibfield  {author} {\bibinfo {author} {\bibfnamefont {D.~S.}\ \bibnamefont
  {Sorenson}}, \bibinfo {author} {\bibfnamefont {G.~A.}\ \bibnamefont
  {Capelle}}, \bibinfo {author} {\bibfnamefont {M.}~\bibnamefont {Grover}},
  \bibinfo {author} {\bibfnamefont {R.~P.}\ \bibnamefont {Johnson}}, \ and\
  \bibinfo {author} {\bibfnamefont {W.~D.}\ \bibnamefont {Turley}},\ }\bibfield
   {title} {\enquote {\bibinfo {title} {Measurements of {Sn} ejecta
  particle-size distributions using ultraviolet in-line fraunhofer
  holography},}\ }\href@noop {} {\bibfield  {journal} {\bibinfo  {journal} {J.
  Dyn. Behav. Mater.}\ }\textbf {\bibinfo {volume} {3}},\ \bibinfo {pages}
  {233} (\bibinfo {year} {2017})}\BibitemShut {NoStop}%
\bibitem [{\citenamefont {Hammerberg}\ \emph {et~al.}(2017)\citenamefont
  {Hammerberg}, \citenamefont {Buttler}, \citenamefont {Llobet}, \citenamefont
  {Morris}, \citenamefont {Goett}, \citenamefont {Manzanares}, \citenamefont
  {Saunders}, \citenamefont {Schmidt}, \citenamefont {Tainter},\ and\
  \citenamefont {Vogan-Mcneil}}]{Hammerberg:1}%
  \BibitemOpen
  \bibfield  {author} {\bibinfo {author} {\bibfnamefont {J.~E.}\ \bibnamefont
  {Hammerberg}}, \bibinfo {author} {\bibfnamefont {W.~T.}\ \bibnamefont
  {Buttler}}, \bibinfo {author} {\bibfnamefont {A.}~\bibnamefont {Llobet}},
  \bibinfo {author} {\bibfnamefont {C.}~\bibnamefont {Morris}}, \bibinfo
  {author} {\bibfnamefont {J.}~\bibnamefont {Goett}}, \bibinfo {author}
  {\bibfnamefont {R.}~\bibnamefont {Manzanares}}, \bibinfo {author}
  {\bibfnamefont {A.}~\bibnamefont {Saunders}}, \bibinfo {author}
  {\bibfnamefont {D.}~\bibnamefont {Schmidt}}, \bibinfo {author} {\bibfnamefont
  {A.}~\bibnamefont {Tainter}}, \ and\ \bibinfo {author} {\bibfnamefont
  {W.}~\bibnamefont {Vogan-Mcneil}},\ }\bibfield  {title} {\enquote {\bibinfo
  {title} {Proton radiography measurements and models of ejecta structure in
  shocked {Sn}},}\ }in\ \href@noop {} {\emph {\bibinfo {booktitle} {20th
  Biennial Conference of the APS Topical Group on Shock Compression of
  Condensed Matter}}}\ (\bibinfo {year} {2017})\BibitemShut {NoStop}%
\bibitem [{\citenamefont {Monfared}\ \emph {et~al.}(2015)\citenamefont
  {Monfared}, \citenamefont {Buttler}, \citenamefont {Frayer}, \citenamefont
  {Grover}, \citenamefont {LaLone}, \citenamefont {Stevens}, \citenamefont
  {Stone}, \citenamefont {Turley},\ and\ \citenamefont {Schauer}}]{Monfared:2}%
  \BibitemOpen
  \bibfield  {author} {\bibinfo {author} {\bibfnamefont {S.~K.}\ \bibnamefont
  {Monfared}}, \bibinfo {author} {\bibfnamefont {W.~T.}\ \bibnamefont
  {Buttler}}, \bibinfo {author} {\bibfnamefont {D.~K.}\ \bibnamefont {Frayer}},
  \bibinfo {author} {\bibfnamefont {M.}~\bibnamefont {Grover}}, \bibinfo
  {author} {\bibfnamefont {B.~M.}\ \bibnamefont {LaLone}}, \bibinfo {author}
  {\bibfnamefont {G.~D.}\ \bibnamefont {Stevens}}, \bibinfo {author}
  {\bibfnamefont {J.~B.}\ \bibnamefont {Stone}}, \bibinfo {author}
  {\bibfnamefont {W.~D.}\ \bibnamefont {Turley}}, \ and\ \bibinfo {author}
  {\bibfnamefont {M.~M.}\ \bibnamefont {Schauer}},\ }\bibfield  {title}
  {\enquote {\bibinfo {title} {Ejected particle size measurement using {Mie}
  scattering in high explosive driven shockwave experiments},}\ }\href
  {\doibase 10.1063/1.4922180} {\bibfield  {journal} {\bibinfo  {journal} {J.
  Appl. Phys.}\ }\textbf {\bibinfo {volume} {117}},\ \bibinfo {pages} {223105}
  (\bibinfo {year} {2015})}\BibitemShut {NoStop}%
\bibitem [{\citenamefont {La-Lone}\ \emph {et~al.}(2015)\citenamefont
  {La-Lone}, \citenamefont {Marshall}, \citenamefont {Miller}, \citenamefont
  {Stevens}, \citenamefont {Turley},\ and\ \citenamefont {Veeser}}]{La:1}%
  \BibitemOpen
  \bibfield  {author} {\bibinfo {author} {\bibfnamefont {B.~M.}\ \bibnamefont
  {La-Lone}}, \bibinfo {author} {\bibfnamefont {B.~R.}\ \bibnamefont
  {Marshall}}, \bibinfo {author} {\bibfnamefont {E.~K.}\ \bibnamefont
  {Miller}}, \bibinfo {author} {\bibfnamefont {G.~D.}\ \bibnamefont {Stevens}},
  \bibinfo {author} {\bibfnamefont {W.~D.}\ \bibnamefont {Turley}}, \ and\
  \bibinfo {author} {\bibfnamefont {L.~R.}\ \bibnamefont {Veeser}},\ }\bibfield
   {title} {\enquote {\bibinfo {title} {Simultaneous broadband laser ranging
  and photonic {D}oppler velocimetry for dynamic compression experiments},}\
  }\href@noop {} {\bibfield  {journal} {\bibinfo  {journal} {Rev. Sci.
  Instrum.}\ }\textbf {\bibinfo {volume} {86}},\ \bibinfo {pages} {4669}
  (\bibinfo {year} {2015})}\BibitemShut {NoStop}%
\bibitem [{\citenamefont {Ogorodnikov}\ \emph {et~al.}(2017)\citenamefont
  {Ogorodnikov}, \citenamefont {Mikhaylov}, \citenamefont {Erunov},
  \citenamefont {Antipov},\ and\ \citenamefont {Chudakov}}]{Ogorodnikov:2}%
  \BibitemOpen
  \bibfield  {author} {\bibinfo {author} {\bibfnamefont {V.~A.}\ \bibnamefont
  {Ogorodnikov}}, \bibinfo {author} {\bibfnamefont {A.~L.}\ \bibnamefont
  {Mikhaylov}}, \bibinfo {author} {\bibfnamefont {S.~V.}\ \bibnamefont
  {Erunov}}, \bibinfo {author} {\bibfnamefont {M.~V.}\ \bibnamefont {Antipov}},
  \ and\ \bibinfo {author} {\bibfnamefont {E.~A.}\ \bibnamefont {Chudakov}},\
  }\bibfield  {title} {\enquote {\bibinfo {title} {Peculiarities of shockwave
  ejecta in the presence of gas in front of a free surface of a material},}\
  }\href@noop {} {\bibfield  {journal} {\bibinfo  {journal} {J. Dyn. Behav.
  Mater.}\ }\textbf {\bibinfo {volume} {3}},\ \bibinfo {pages} {225--232}
  (\bibinfo {year} {2017})}\BibitemShut {NoStop}%
\bibitem [{\citenamefont {Andriyash}\ \emph {et~al.}(2018)\citenamefont
  {Andriyash}, \citenamefont {Astashkin}, \citenamefont {Baranov},
  \citenamefont {Golubinskii}, \citenamefont {Irinichev}, \citenamefont
  {Khatunkin}, \citenamefont {Kondratev}, \citenamefont {Kuratov},
  \citenamefont {Mazanov}, \citenamefont {Rogozkin},\ and\ \citenamefont
  {Stepushkin}}]{Andriyash:1}%
  \BibitemOpen
  \bibfield  {author} {\bibinfo {author} {\bibfnamefont {A.~V.}\ \bibnamefont
  {Andriyash}}, \bibinfo {author} {\bibfnamefont {M.~V.}\ \bibnamefont
  {Astashkin}}, \bibinfo {author} {\bibfnamefont {V.~K.}\ \bibnamefont
  {Baranov}}, \bibinfo {author} {\bibfnamefont {A.~G.}\ \bibnamefont
  {Golubinskii}}, \bibinfo {author} {\bibfnamefont {D.~A.}\ \bibnamefont
  {Irinichev}}, \bibinfo {author} {\bibfnamefont {V.~Y.}\ \bibnamefont
  {Khatunkin}}, \bibinfo {author} {\bibfnamefont {A.~N.}\ \bibnamefont
  {Kondratev}}, \bibinfo {author} {\bibfnamefont {S.~E.}\ \bibnamefont
  {Kuratov}}, \bibinfo {author} {\bibfnamefont {V.~A.}\ \bibnamefont
  {Mazanov}}, \bibinfo {author} {\bibfnamefont {D.~B.}\ \bibnamefont
  {Rogozkin}}, \ and\ \bibinfo {author} {\bibfnamefont {S.~N.}\ \bibnamefont
  {Stepushkin}},\ }\bibfield  {title} {\enquote {\bibinfo {title} {Application
  of photon doppler velocimetry for characterization of ejecta from
  shock-loaded samples},}\ }\href {\doibase 10.1063/1.5029958} {\bibfield
  {journal} {\bibinfo  {journal} {J. Appl. Phys.}\ }\textbf {\bibinfo {volume}
  {123}},\ \bibinfo {pages} {243102} (\bibinfo {year} {2018})}\BibitemShut
  {NoStop}%
\bibitem [{\citenamefont {Franzkowiak}\ \emph
  {et~al.}(2018{\natexlab{a}})\citenamefont {Franzkowiak}, \citenamefont
  {Prudhomme}, \citenamefont {Mercier}, \citenamefont {Lauriot}, \citenamefont
  {Dubreuil},\ and\ \citenamefont {Berthe}}]{Franzkowiak:1}%
  \BibitemOpen
  \bibfield  {author} {\bibinfo {author} {\bibfnamefont {J.~E.}\ \bibnamefont
  {Franzkowiak}}, \bibinfo {author} {\bibfnamefont {G.}~\bibnamefont
  {Prudhomme}}, \bibinfo {author} {\bibfnamefont {P.}~\bibnamefont {Mercier}},
  \bibinfo {author} {\bibfnamefont {S.}~\bibnamefont {Lauriot}}, \bibinfo
  {author} {\bibfnamefont {E.}~\bibnamefont {Dubreuil}}, \ and\ \bibinfo
  {author} {\bibfnamefont {L.}~\bibnamefont {Berthe}},\ }\bibfield  {title}
  {\enquote {\bibinfo {title} {{PDV}-based estimation of ejecta particles'
  mass-velocity function from shock-loaded tin experiment},}\ }\href {\doibase
  10.1002/cphc.201700515} {\bibfield  {journal} {\bibinfo  {journal} {Rev. Sci.
  Instrum.}\ }\textbf {\bibinfo {volume} {89}},\ \bibinfo {pages} {033901}
  (\bibinfo {year} {2018}{\natexlab{a}})}\BibitemShut {NoStop}%
\bibitem [{\citenamefont {Sun}\ \emph {et~al.}(2016)\citenamefont {Sun},
  \citenamefont {Wang}, \citenamefont {Chen},\ and\ \citenamefont
  {Ma}}]{Sun:1}%
  \BibitemOpen
  \bibfield  {author} {\bibinfo {author} {\bibfnamefont {H.}~\bibnamefont
  {Sun}}, \bibinfo {author} {\bibfnamefont {P.}~\bibnamefont {Wang}}, \bibinfo
  {author} {\bibfnamefont {D.}~\bibnamefont {Chen}}, \ and\ \bibinfo {author}
  {\bibfnamefont {D.}~\bibnamefont {Ma}},\ }\bibfield  {title} {\enquote
  {\bibinfo {title} {A new method to analyze the velocity spectrograms of
  photonic {D}oppler velocimetry},}\ }\href {\doibase 10.7498/aps.65.104702}
  {\bibfield  {journal} {\bibinfo  {journal} {Acta Physica Sinica}\ }\textbf
  {\bibinfo {volume} {65}},\ \bibinfo {pages} {104702} (\bibinfo {year}
  {2016})}\BibitemShut {NoStop}%
\bibitem [{\citenamefont {Fedorov}, \citenamefont {Gnutov},\ and\ \citenamefont
  {Yagovkin}(2018)}]{Fedorov:1}%
  \BibitemOpen
  \bibfield  {author} {\bibinfo {author} {\bibfnamefont {A.~V.}\ \bibnamefont
  {Fedorov}}, \bibinfo {author} {\bibfnamefont {I.~S.}\ \bibnamefont {Gnutov}},
  \ and\ \bibinfo {author} {\bibfnamefont {A.~O.}\ \bibnamefont {Yagovkin}},\
  }\bibfield  {title} {\enquote {\bibinfo {title} {Determination of the sizes
  of particle ejected from shock-loaded surfaces during their deceleration in a
  gaseous medium},}\ }\href {\doibase 10.1134/S1063776118010156} {\bibfield
  {journal} {\bibinfo  {journal} {J. Exp. Theor. Phys.}\ }\textbf {\bibinfo
  {volume} {126}},\ \bibinfo {pages} {76} (\bibinfo {year} {2018})}\BibitemShut
  {NoStop}%
\bibitem [{\citenamefont {Kondrat'Ev}, \citenamefont {Andriyash},\ and\
  \citenamefont {Kuratov}(2020)}]{Arsenii:1}%
  \BibitemOpen
  \bibfield  {author} {\bibinfo {author} {\bibfnamefont {A.}~\bibnamefont
  {Kondrat'Ev}}, \bibinfo {author} {\bibfnamefont {A.~V.}\ \bibnamefont
  {Andriyash}}, \ and\ \bibinfo {author} {\bibfnamefont {S.~E.}\ \bibnamefont
  {Kuratov}},\ }\bibfield  {title} {\enquote {\bibinfo {title} {Application of
  multiple scattering theory to {D}oppler velocimetry of ejecta from
  shock-loaded samples},}\ }\href {\doibase 10.1016/j.jqsrt.2020.106925}
  {\bibfield  {journal} {\bibinfo  {journal} {J. Quant. Spectrosc. Radiat.
  Transf.}\ }\textbf {\bibinfo {volume} {246}},\ \bibinfo {pages} {106925}
  (\bibinfo {year} {2020})}\BibitemShut {NoStop}%
\bibitem [{\citenamefont {Buttler}\ \emph {et~al.}(2009)\citenamefont
  {Buttler}, \citenamefont {Oro}, \citenamefont {Dimonte}, \citenamefont
  {Terrones}, \citenamefont {Morris}, \citenamefont {Bainbridge}, \citenamefont
  {Hogan}, \citenamefont {Hollander}, \citenamefont {Holtkamp}, \citenamefont
  {Kwiathowski}, \citenamefont {Marr-Lyon}, \citenamefont {Mariam},
  \citenamefont {Merrill}, \citenamefont {Nedrow}, \citenamefont {Saunders},
  \citenamefont {Schwartz}, \citenamefont {Stone}, \citenamefont {Tupa},\ and\
  \citenamefont {Vogan-McNeil}}]{Buttler:1}%
  \BibitemOpen
  \bibfield  {author} {\bibinfo {author} {\bibfnamefont {W.~T.}\ \bibnamefont
  {Buttler}}, \bibinfo {author} {\bibfnamefont {D.~M.}\ \bibnamefont {Oro}},
  \bibinfo {author} {\bibfnamefont {G.}~\bibnamefont {Dimonte}}, \bibinfo
  {author} {\bibfnamefont {G.}~\bibnamefont {Terrones}}, \bibinfo {author}
  {\bibfnamefont {C.}~\bibnamefont {Morris}}, \bibinfo {author} {\bibfnamefont
  {J.~R.}\ \bibnamefont {Bainbridge}}, \bibinfo {author} {\bibfnamefont
  {G.~E.}\ \bibnamefont {Hogan}}, \bibinfo {author} {\bibfnamefont {B.~J.}\
  \bibnamefont {Hollander}}, \bibinfo {author} {\bibfnamefont {D.~B.}\
  \bibnamefont {Holtkamp}}, \bibinfo {author} {\bibfnamefont {K.}~\bibnamefont
  {Kwiathowski}}, \bibinfo {author} {\bibfnamefont {M.}~\bibnamefont
  {Marr-Lyon}}, \bibinfo {author} {\bibfnamefont {F.~G.}\ \bibnamefont
  {Mariam}}, \bibinfo {author} {\bibfnamefont {F.~E.}\ \bibnamefont {Merrill}},
  \bibinfo {author} {\bibfnamefont {P.}~\bibnamefont {Nedrow}}, \bibinfo
  {author} {\bibfnamefont {A.}~\bibnamefont {Saunders}}, \bibinfo {author}
  {\bibfnamefont {C.~L.}\ \bibnamefont {Schwartz}}, \bibinfo {author}
  {\bibfnamefont {B.}~\bibnamefont {Stone}}, \bibinfo {author} {\bibfnamefont
  {D.}~\bibnamefont {Tupa}}, \ and\ \bibinfo {author} {\bibfnamefont {W.~S.}\
  \bibnamefont {Vogan-McNeil}},\ }\bibfield  {title} {\enquote {\bibinfo
  {title} {Ejecta model development at p{R}ad (u)},}\ }in\ \href@noop {} {\emph
  {\bibinfo {booktitle} {Proceedings NEDPC 2009}}},\ \bibinfo {series and
  number} {\bibinfo {number} {LA-UR-10-00734}}\ (\bibinfo {year}
  {2009})\BibitemShut {NoStop}%
\bibitem [{\citenamefont {Buttler}\ \emph {et~al.}(2012)\citenamefont
  {Buttler}, \citenamefont {Or{\'o}}, \citenamefont {Preston}, \citenamefont
  {Mikaelian}, \citenamefont {Cherne}, \citenamefont {Hixson}, \citenamefont
  {Mariam}, \citenamefont {Morris}, \citenamefont {Stone}, \citenamefont
  {Terrones},\ and\ \citenamefont {Tupa}}]{Buttler:2}%
  \BibitemOpen
  \bibfield  {author} {\bibinfo {author} {\bibfnamefont {W.~T.}\ \bibnamefont
  {Buttler}}, \bibinfo {author} {\bibfnamefont {D.~M.}\ \bibnamefont
  {Or{\'o}}}, \bibinfo {author} {\bibfnamefont {D.~L.}\ \bibnamefont
  {Preston}}, \bibinfo {author} {\bibfnamefont {K.~O.}\ \bibnamefont
  {Mikaelian}}, \bibinfo {author} {\bibfnamefont {F.~J.}\ \bibnamefont
  {Cherne}}, \bibinfo {author} {\bibfnamefont {R.~S.}\ \bibnamefont {Hixson}},
  \bibinfo {author} {\bibfnamefont {F.~G.}\ \bibnamefont {Mariam}}, \bibinfo
  {author} {\bibfnamefont {C.}~\bibnamefont {Morris}}, \bibinfo {author}
  {\bibfnamefont {J.~B.}\ \bibnamefont {Stone}}, \bibinfo {author}
  {\bibfnamefont {G.}~\bibnamefont {Terrones}}, \ and\ \bibinfo {author}
  {\bibfnamefont {D.}~\bibnamefont {Tupa}},\ }\bibfield  {title} {\enquote
  {\bibinfo {title} {Unstable richtmyer–meshkov growth of solid and liquid
  metals in vacuum},}\ }\href@noop {} {\bibfield  {journal} {\bibinfo
  {journal} {J. Fluid Mech.}\ }\textbf {\bibinfo {volume} {703}},\ \bibinfo
  {pages} {60--87} (\bibinfo {year} {2012})}\BibitemShut {NoStop}%
\bibitem [{\citenamefont {Ishimaru}(1978)}]{Ishimaru:1}%
  \BibitemOpen
  \bibfield  {author} {\bibinfo {author} {\bibfnamefont {A.}~\bibnamefont
  {Ishimaru}},\ }\href@noop {} {\emph {\bibinfo {title} {Wave propagation and
  scattering in random media}}}\ (\bibinfo  {publisher} {Academic Press},\
  \bibinfo {year} {1978})\BibitemShut {NoStop}%
\bibitem [{\citenamefont {Reguigui}\ \emph {et~al.}(1997)\citenamefont
  {Reguigui}, \citenamefont {Ackerson}, \citenamefont {Dorri-Nowkoorani},
  \citenamefont {Dougherty},\ and\ \citenamefont {Nobbmann}}]{Reguigui:1}%
  \BibitemOpen
  \bibfield  {author} {\bibinfo {author} {\bibfnamefont {N.~M.}\ \bibnamefont
  {Reguigui}}, \bibinfo {author} {\bibfnamefont {B.~J.}\ \bibnamefont
  {Ackerson}}, \bibinfo {author} {\bibfnamefont {F.}~\bibnamefont
  {Dorri-Nowkoorani}}, \bibinfo {author} {\bibfnamefont {R.~L.}\ \bibnamefont
  {Dougherty}}, \ and\ \bibinfo {author} {\bibfnamefont {U.}~\bibnamefont
  {Nobbmann}},\ }\bibfield  {title} {\enquote {\bibinfo {title} {Correlation
  transfer: Index of refraction and anisotropy effects},}\ }\href@noop {}
  {\bibfield  {journal} {\bibinfo  {journal} {J. Thermophys. Heat Transf.}\
  }\textbf {\bibinfo {volume} {11}},\ \bibinfo {pages} {400} (\bibinfo {year}
  {1997})}\BibitemShut {NoStop}%
\bibitem [{\citenamefont {Binzoni}\ \emph {et~al.}(2016)\citenamefont
  {Binzoni}, \citenamefont {Liemert}, \citenamefont {Kienle},\ and\
  \citenamefont {Martelli}}]{Binzoni:1}%
  \BibitemOpen
  \bibfield  {author} {\bibinfo {author} {\bibfnamefont {T.}~\bibnamefont
  {Binzoni}}, \bibinfo {author} {\bibfnamefont {A.}~\bibnamefont {Liemert}},
  \bibinfo {author} {\bibfnamefont {A.}~\bibnamefont {Kienle}}, \ and\ \bibinfo
  {author} {\bibfnamefont {F.}~\bibnamefont {Martelli}},\ }\bibfield  {title}
  {\enquote {\bibinfo {title} {Analytical solution of the correlation transport
  equation with static background: Beyond diffuse correlation spectroscopy},}\
  }\href@noop {} {\bibfield  {journal} {\bibinfo  {journal} {Appl. Opt.}\
  }\textbf {\bibinfo {volume} {55}},\ \bibinfo {pages} {8500} (\bibinfo {year}
  {2016})}\BibitemShut {NoStop}%
\bibitem [{\citenamefont {Bohren}\ and\ \citenamefont
  {Huffman}(2004)}]{Bohren:1}%
  \BibitemOpen
  \bibfield  {author} {\bibinfo {author} {\bibfnamefont {C.~F.}\ \bibnamefont
  {Bohren}}\ and\ \bibinfo {author} {\bibfnamefont {D.~R.}\ \bibnamefont
  {Huffman}},\ }\href@noop {} {\emph {\bibinfo {title} {Absorption and
  Scattering of Light by Small Particles}}}\ (\bibinfo  {publisher} {Wiley-VCH
  Verlag GmbH \& Co. KGaA},\ \bibinfo {year} {2004})\BibitemShut {NoStop}%
\bibitem [{\citenamefont {Mie}(1908)}]{Mie:1}%
  \BibitemOpen
  \bibfield  {author} {\bibinfo {author} {\bibfnamefont {G.}~\bibnamefont
  {Mie}},\ }\bibfield  {title} {\enquote {\bibinfo {title} {Contributions to
  the optics of turbid media, particularly of colloidal metal solutions},}\
  }\href@noop {} {\bibfield  {journal} {\bibinfo  {journal} {Ann. Phys.}\
  }\textbf {\bibinfo {volume} {330}} (\bibinfo {year} {1908})}\BibitemShut
  {NoStop}%
\bibitem [{\citenamefont {Durand}\ and\ \citenamefont
  {Soulard}(2015)}]{Durand:1}%
  \BibitemOpen
  \bibfield  {author} {\bibinfo {author} {\bibfnamefont {O.}~\bibnamefont
  {Durand}}\ and\ \bibinfo {author} {\bibfnamefont {L.}~\bibnamefont
  {Soulard}},\ }\bibfield  {title} {\enquote {\bibinfo {title} {Mass-velocity
  and size-velocity distributions of ejecta cloud from shock-loaded tin surface
  using atomistic simulations},}\ }\href@noop {} {\bibfield  {journal}
  {\bibinfo  {journal} {J. Appl. Phys.}\ }\textbf {\bibinfo {volume} {117}},\
  \bibinfo {pages} {024905--797} (\bibinfo {year} {2015})}\BibitemShut
  {NoStop}%
\bibitem [{\citenamefont {Durand}\ and\ \citenamefont
  {Soulard}(2012)}]{Durand:2}%
  \BibitemOpen
  \bibfield  {author} {\bibinfo {author} {\bibfnamefont {O.}~\bibnamefont
  {Durand}}\ and\ \bibinfo {author} {\bibfnamefont {L.}~\bibnamefont
  {Soulard}},\ }\bibfield  {title} {\enquote {\bibinfo {title} {Large-scale
  molecular dynamics study of jet breakup and ejecta production from
  shock-loaded copper with a hybrid method},}\ }\href@noop {} {\bibfield
  {journal} {\bibinfo  {journal} {J. Appl. Phys.}\ }\textbf {\bibinfo {volume}
  {111}},\ \bibinfo {pages} {284} (\bibinfo {year} {2012})}\BibitemShut
  {NoStop}%
\bibitem [{\citenamefont {Schauer}\ \emph {et~al.}(2017)\citenamefont
  {Schauer}, \citenamefont {Buttler}, \citenamefont {Frayer}, \citenamefont
  {Grover},\ and\ \citenamefont {Turley}}]{Schauer:1}%
  \BibitemOpen
  \bibfield  {author} {\bibinfo {author} {\bibfnamefont {M.~M.}\ \bibnamefont
  {Schauer}}, \bibinfo {author} {\bibfnamefont {W.~T.}\ \bibnamefont
  {Buttler}}, \bibinfo {author} {\bibfnamefont {D.~K.}\ \bibnamefont {Frayer}},
  \bibinfo {author} {\bibfnamefont {M.}~\bibnamefont {Grover}}, \ and\ \bibinfo
  {author} {\bibfnamefont {W.~D.}\ \bibnamefont {Turley}},\ }\bibfield  {title}
  {\enquote {\bibinfo {title} {Ejected particle size distributions from shocked
  metal surfaces},}\ }\href@noop {} {\bibfield  {journal} {\bibinfo  {journal}
  {J. Dyn. Behav. Mater.}\ }\textbf {\bibinfo {volume} {3}},\ \bibinfo {pages}
  {217} (\bibinfo {year} {2017})}\BibitemShut {NoStop}%
\bibitem [{\citenamefont {He}\ \emph {et~al.}(2017)\citenamefont {He},
  \citenamefont {Wang}, \citenamefont {Shao},\ and\ \citenamefont
  {Duan}}]{He:2}%
  \BibitemOpen
  \bibfield  {author} {\bibinfo {author} {\bibfnamefont {A.}~\bibnamefont
  {He}}, \bibinfo {author} {\bibfnamefont {P.}~\bibnamefont {Wang}}, \bibinfo
  {author} {\bibfnamefont {J.}~\bibnamefont {Shao}}, \ and\ \bibinfo {author}
  {\bibfnamefont {S.}~\bibnamefont {Duan}},\ }\bibfield  {title} {\enquote
  {\bibinfo {title} {Molecular dynamics simulations of jet breakup and ejecta
  production from a grooved {Cu} surface under shock loading},}\ }\href@noop {}
  {\bibfield  {journal} {\bibinfo  {journal} {Chin. Phys. B}\ }\textbf
  {\bibinfo {volume} {23}},\ \bibinfo {pages} {047102} (\bibinfo {year}
  {2017})}\BibitemShut {NoStop}%
\bibitem [{\citenamefont {Bell}\ \emph {et~al.}(2017)\citenamefont {Bell},
  \citenamefont {Routley}, \citenamefont {Millett}, \citenamefont {Whiteman},\
  and\ \citenamefont {Keightley}}]{Bell:1}%
  \BibitemOpen
  \bibfield  {author} {\bibinfo {author} {\bibfnamefont {D.~J.}\ \bibnamefont
  {Bell}}, \bibinfo {author} {\bibfnamefont {N.~R.}\ \bibnamefont {Routley}},
  \bibinfo {author} {\bibfnamefont {J.~C.~F.}\ \bibnamefont {Millett}},
  \bibinfo {author} {\bibfnamefont {G.}~\bibnamefont {Whiteman}}, \ and\
  \bibinfo {author} {\bibfnamefont {P.~T.}\ \bibnamefont {Keightley}},\
  }\bibfield  {title} {\enquote {\bibinfo {title} {Investigation of ejecta
  production from tin at an elevated temperature and the eutectic alloy
  lead-bismuth},}\ }\href@noop {} {\bibfield  {journal} {\bibinfo  {journal}
  {J. Dyn. Behav. Mater.}\ }\textbf {\bibinfo {volume} {3}},\ \bibinfo {pages}
  {208} (\bibinfo {year} {2017})}\BibitemShut {NoStop}%
\bibitem [{\citenamefont {Andriyash}\ \emph {et~al.}(2020)\citenamefont
  {Andriyash}, \citenamefont {Dyachkova}, \citenamefont {Zhakhovskya},
  \citenamefont {Kalashnikovc}, \citenamefont {Kondrateva}, \citenamefont
  {Kuratova}, \citenamefont {Mikhailovc}, \citenamefont {Rogozkina},
  \citenamefont {Fedorovc}, \citenamefont {Finyushinc},\ and\ \citenamefont
  {Chudakovc}}]{Andriyash:2}%
  \BibitemOpen
  \bibfield  {author} {\bibinfo {author} {\bibfnamefont {A.~V.}\ \bibnamefont
  {Andriyash}}, \bibinfo {author} {\bibfnamefont {S.~A.}\ \bibnamefont
  {Dyachkova}}, \bibinfo {author} {\bibfnamefont {V.~V.}\ \bibnamefont
  {Zhakhovskya}}, \bibinfo {author} {\bibfnamefont {D.~A.}\ \bibnamefont
  {Kalashnikovc}}, \bibinfo {author} {\bibfnamefont {A.~N.}\ \bibnamefont
  {Kondrateva}}, \bibinfo {author} {\bibfnamefont {S.~E.}\ \bibnamefont
  {Kuratova}}, \bibinfo {author} {\bibfnamefont {A.~L.}\ \bibnamefont
  {Mikhailovc}}, \bibinfo {author} {\bibfnamefont {D.~B.}\ \bibnamefont
  {Rogozkina}}, \bibinfo {author} {\bibfnamefont {A.~V.}\ \bibnamefont
  {Fedorovc}}, \bibinfo {author} {\bibfnamefont {S.~A.}\ \bibnamefont
  {Finyushinc}}, \ and\ \bibinfo {author} {\bibfnamefont {E.~A.}\ \bibnamefont
  {Chudakovc}},\ }\bibfield  {title} {\enquote {\bibinfo {title} {Photon
  doppler velocimetry and simulation of ejection of particles from the surface
  of shock-loaded samples},}\ }\href@noop {} {\bibfield  {journal} {\bibinfo
  {journal} {J. Exp. Theor. Phys.}\ }\textbf {\bibinfo {volume} {130}},\
  \bibinfo {pages} {338} (\bibinfo {year} {2020})}\BibitemShut {NoStop}%
\bibitem [{\citenamefont {Franzkowiak}\ \emph
  {et~al.}(2018{\natexlab{b}})\citenamefont {Franzkowiak}, \citenamefont
  {Mercier}, \citenamefont {Prudhomme},\ and\ \citenamefont
  {Berthe}}]{Franzkowiak:2}%
  \BibitemOpen
  \bibfield  {author} {\bibinfo {author} {\bibfnamefont {J.~E.}\ \bibnamefont
  {Franzkowiak}}, \bibinfo {author} {\bibfnamefont {P.}~\bibnamefont
  {Mercier}}, \bibinfo {author} {\bibfnamefont {G.}~\bibnamefont {Prudhomme}},
  \ and\ \bibinfo {author} {\bibfnamefont {L.}~\bibnamefont {Berthe}},\
  }\bibfield  {title} {\enquote {\bibinfo {title} {Multiple light scattering in
  metallic ejecta produced under intense shockwave compression},}\ }\href@noop
  {} {\bibfield  {journal} {\bibinfo  {journal} {Appl. Opt.}\ }\textbf
  {\bibinfo {volume} {57}},\ \bibinfo {pages} {2766} (\bibinfo {year}
  {2018}{\natexlab{b}})}\BibitemShut {NoStop}%
\bibitem [{\citenamefont {Walsh}, \citenamefont {Shreffler},\ and\
  \citenamefont {Willig}(1953)}]{Walsh:1}%
  \BibitemOpen
  \bibfield  {author} {\bibinfo {author} {\bibfnamefont {J.~M.}\ \bibnamefont
  {Walsh}}, \bibinfo {author} {\bibfnamefont {R.~G.}\ \bibnamefont
  {Shreffler}}, \ and\ \bibinfo {author} {\bibfnamefont {F.~J.}\ \bibnamefont
  {Willig}},\ }\bibfield  {title} {\enquote {\bibinfo {title} {Limiting
  conditions for jet formation in high velocity collisions},}\ }\href {\doibase
  10.1063/1.1721278} {\bibfield  {journal} {\bibinfo  {journal} {J. Appl.
  Phys.}\ }\textbf {\bibinfo {volume} {24}},\ \bibinfo {pages} {349} (\bibinfo
  {year} {1953})}\BibitemShut {NoStop}%
\bibitem [{\citenamefont {Mader}(1980)}]{Mader:1}%
  \BibitemOpen
  \bibfield  {author} {\bibinfo {author} {\bibfnamefont {C.}~\bibnamefont
  {Mader}},\ }\href@noop {} {\emph {\bibinfo {title} {LASL PHERMEX Data,
  Volumes I, II, III}}}\ (\bibinfo  {publisher} {University of California
  Press},\ \bibinfo {address} {Berkeley},\ \bibinfo {year} {1980})\BibitemShut
  {NoStop}%
\end{thebibliography}%

\end{document}